\documentclass[11pt,journal,draftcls,onecolumn,peerreviewca]{IEEEtran}
\usepackage{amsfonts}
\usepackage{amsmath}
\usepackage{amssymb}
\usepackage{bm}
\usepackage{xcolor,graphicx,float}
\usepackage{color, soul}
\usepackage{stfloats}
\usepackage[numbers,sort&compress]{natbib}
\usepackage[amsmath,thmmarks]{ntheorem}
\usepackage{theorem}
\usepackage{caption}
\usepackage{algorithm}
\usepackage{algorithmic}
\usepackage{graphicx}
\usepackage{subfigure}
\usepackage{pict2e}
\usepackage{comment}
\usepackage{makecell}

\theoremheaderfont{\sc}\theorembodyfont{\upshape}
\theoremstyle{nonumberplain}
\theoremseparator{}
\theoremsymbol{\rule{1ex}{1ex}}

\begin{document}
	\title{A Model and Data Dual-driven Approach for Multitargets Detection under Mainlobe Jamming}
	\author{Ruohai Guo, Jiang Zhu, Chengjie Yu, Zhigang Wang, Ning Zhang, Fengzhong Qu and Min Gong
\thanks{(\emph{Corresponding Author: Jiang Zhu.}).}}
	\maketitle
	
	\begin{abstract}
In modern radar systems, target detection and parameter estimation face significant challenges when confronted with  mainlobe jamming. This paper presents a Diffusion-based Model and Data Dual-driven (DMDD) approach to  estimate and detect multitargets and suppress structured jamming. In DMDD, the jamming prior is modeled through a score-based diffusion process with its score learned from the pure jamming data, enabling posterior sampling without requiring detailed knowledge of jamming. Meanwhile, the target signal is usually sparse in the range space, which can be modeled via a sparse Bayesian learning (SBL) framework, and hyperparameter is updated through the expectation-maximization (EM) algorithm. A single diffusion process is constructed for the jamming, while the state of targets are estimated through direct posterior inference, enhancing computational efficiency. The noise variance is also estimated through EM algorithm. Numerical experiments demonstrate the effectiveness of the proposed method in structured jamming scenarios. The proposed DMDD algorithm achieves superior target detection performance, compared with existing methods.
	\end{abstract}
	
	\section{Introduction}
For modern radar systems, mainlobe jamming (MLJ) is one of the most critical threats, as it directly contaminates the receiver’s main beam, significantly degrading the performance of multitargets detection and parameter estimation. Unlike sidelobe jamming, which can often be suppressed through adaptive sidelobe cancellation (ASLC) \cite{GSC,AGSC,LCMV} or sidelobe blanking (SLB) algorithms \cite{Sideblank,Sideblank2}, MLJ overlaps with the angular region of interest \cite{antijam}. This makes it particularly challenging to suppress without distorting the desired target signals \cite{DOA}. As a result, reliable detection of multitargets under strong MLJ remain unresolved issues with high practical importance in radar and electronic warfare systems. 

Filtering methods aim to spatially nullify the jamming, such as adaptive beamforming \cite{jointBF, BF2, BF3}, and space-time adaptive processing (STAP) \cite{STAP,STAP2}. Among the techniques proposed, Chen et al. introduced an adaptive algorithm that leverages multi-radar joint beamspace processing to mitigate multi-mainlobe jamming \cite{jointBF}. Blocking matrix processing beamforming (BMPB) relies on the cancellation of jamming through the use of adjacent antennas \cite{BMPB}. Eigen-projection matrix preprocessing (EMP) focuses on extracting eigenvalues from the signal’s covariance matrix to enhance the signal quality \cite{EMP}. While effective in certain cases, these methods often suffer performance loss when the jamming signal is aligned with the mainlobe, and may distort the beam, resulting in a loss of signal-to-noise ratio (SNR). 

Recently, diffusion models (DMs) have emerged as a powerful class of generative methods capable of modeling complex, high-dimensional distributions through iterative denoising processes \cite{SongICLR,Ho}. Originally proposed for image synthesis and restoration, these models have demonstrated strong performance in tasks that require sample generation from intractable posteriors \cite{Songmed,image,Chung}. To address linear inverse problems under structured jamming, \cite{Steven2025} and \cite{Steven2024} propose learning score models for both the noise and the desired signal to remove structured noise from natural images and ultrasound data, which conduct simultaneously signal and jamming estimation via conditional sampling. In a similar vein, \cite{SBSP} employs score-based diffusion to separate the signal of interest from multiple independent sources, followed by data demodulation. In \cite{yifan}, diffusion model based sparse Bayesian learning (DM-SBL) approach was introduced for underwater acoustic channel estimation under structured jamming, where two independent diffusion model were implemented.
It is expected that in radar signal processing, DMs offer a promising avenue for modeling structured jamming and enabling flexible posterior inference.

In this paper, we address the problem of target estimation and detection in modern radar systems under structured jamming by proposing a model and data dual-driven approach, which is named as Diffusion-based Model and Data Dual-driven (DMDD). It is also worth noting the proposed DMDD could suppress the structured clutter, as evidenced in Section \ref{real}. A data-driven method is employed to characterize the jamming through score function training, enabling generating high fidelity jamming samples. For target detection, a model-based strategy leveraging SBL to exploit the inherent sparsity of the targets is developed. At each iteration, the jamming is first estimated, followed by the posterior inference of the target amplitudes. The hyperparameters governing the sparsity prior and the noise variance are then updated via the expectation-maximization (EM) algorithm. In summary, the main contributions can be summarized as follows.
\begin{enumerate}
    \item A model and data dual-driven approach is proposed to address the target detection under structured mainlobe jamming in modern radar systems. The scenario considered is a stringent case where the jamming overlaps almost completely with the mainlobe, both in the frequency and time domain, making it impossible to suppress the jamming through spatial domain filtering.
    \item The sparsity of the targets is exploited by using SBL, enabling efficient detection of the targets. Posterior inference is performed at each step, and the noise variance  is iteratively estimated using the EM algorithm.
    \item The proposed method demonstrates improved performance in terms of target detection, compared to traditional and state-of-art approaches.
\end{enumerate}

The rest of this paper is organized as follows. In Section \ref{SignalModel}, the signal model is introduced. Section \ref{pretrain} introduces the pre-trained neural network for jamming. Section \ref{method} offers a detailed implementation of the proposed DMDD approach. Section \ref{num} presents the numerical simulation results. 
Finally, Section \ref{con} concludes the paper.

\textit{Notation}: The boldfaced letters $\mathbf{x}, \mathbf{X}$ denote vectors and matrices, respectively. $\mathbf{x}(i)$ denotes the $i$th element of vector $\mathbf{x}$. $(\cdot)^{\mathrm{T}},(\cdot)^{\mathrm{H}}$ and $(\cdot)^*$ represent the transpose, Hermitian and conjugate, respectively. 
$\Re\{\cdot\}$ and $\Im\{\cdot\}$ denote the real and imaginary parts. 
We use $\nabla_{\mathbf{z}^*} f\left(\mathbf{z}, \mathbf{z}^*\right)$ to denote the gradient of a function $f$ with respect to the vector $\mathbf{z}^*$. $\mathcal{N}(\mathbf{x} ; \boldsymbol{\mu}, \boldsymbol{\Sigma})$ and $\mathcal{C} \mathcal{N}(\mathbf{x} ; \boldsymbol{\mu}, \boldsymbol{\Sigma})$ denote the Gaussian distribution and complex Gaussian distribution for random variable $\mathbf{x}$ with mean $\boldsymbol{\mu}$ and covariance $\boldsymbol{\Sigma}$. For a matrix $\boldsymbol{\Sigma}$, $|\boldsymbol{\Sigma}|$ denote its determinant. For a vector $\mathbf{x},|\mathbf{x}|$ and $\mathbf{x}^{\odot 2}$ denotes its elementwise modulus and square, respectively.
	
\section{Signal Model}\label{SignalModel}
\begin{figure}
	\centering
	\includegraphics[width = 6in]{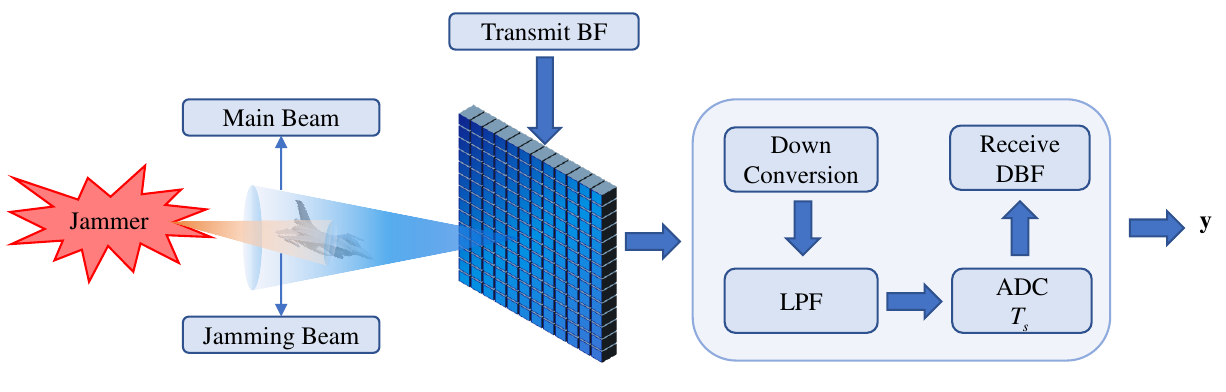}
	\caption{Processing pipeline of the phased array radar.}\label{rxprocess}
\end{figure}

Consider a typical phased array radar system. The transmit antenna applies digital beamforming to transmit the pulse signal to steer a beam in a coherent processing interval. In the receiver side, the signal undergoes a series of standard operations, including down conversion, low pass filtering (LPF), analog to digital conversion (ADC) for sampling and receive digital beamforming (DBF). In the presence of mainlobe jamming, these operations collectively form the receiver processing pipeline, as illustrated in Fig. \ref{rxprocess}, and the corresponding received signal can be described as 
\begin{align}\label{signalmodel} 
    \mathbf{y} = \sum_{k=1}^{K} x_k\mathbf{s}_{N}(r_k)+\mathbf{i}+\mathbf{w},
    \end{align}
where
    \begin{align}
    \mathbf{s}_{N}(r) = \left[s\left(0 T_s-\frac{2 r}{c}\right), s\left(1 T_s-\frac{2 r}{c}\right), \cdots, s\left((N-1) T_s-\frac{2 r}{c}\right)\right]^{\rm{T}},\notag
    \end{align}
$s(t)$ represents the baseband signal which can be any arbitrary waveform with an analytic expression; $N$ denotes the number of fast time domain samples; 
$K$ is the number of targets; $r_k$ denotes the radial distance of the $k$th target, and $x_k$ is the corresponding complex amplitude; $T_s$ denotes the sampling interval; $c$ denotes the light speed.
$\mathbf{i}\in \mathbb{C}^{N \times 1}$ is the mainlobe jamming.
$\mathbf{w}\in \mathbb{C}^{N \times 1}$ is the additive noise and is supposed to be independent and identically distributed (i.i.d.), and the $n$th element ${w}_n \sim {\mathcal {CN}}(0,\sigma_{w}^2)$ with $\sigma_{w}^2$ being the variance of the noise.

By defining 
\begin{align}
    \mathbf{A}(\mathbf{r}) = \left[\mathbf{s}_{N}(r_1), \mathbf{s}_{N}(r_2),\cdots,\mathbf{s}_{N}(r_K) \right]\in \mathbb{C}^{N\times K},
\end{align}
model (\ref{signalmodel}) can be reformulated as
\begin{align}\label{measvec}
    \mathbf{y} = \mathbf{A}(\mathbf{r})\mathbf{x} + \mathbf{i} + \mathbf{w},
\end{align}
where $\mathbf{x}=[x_1,x_2,\cdots,x_K]\in \mathbb{C}^{K\times 1}$. Note that the jamming $\mathbf{i}\sim p(\mathbf{i})$, and the exact expression of $p(\mathbf{i})$ is unknown. However, as stated in Section \ref{pretrain}, the score $\nabla_{{\mathbf i}^*}\log p(\mathbf{i})$ of $\mathbf{i}$, i.e., the Wirtinger calculus of $\log p(\mathbf{i})$ with respect to the conjugate of $\mathbf i$, can be learned from the pure jamming data by training a neural network.

Note that $\mathbf{A}(\mathbf{r})$ in (\ref{measvec}) is nonlinearly coupled with the range parameter $\mathbf{r}$. To simplify the nonlinear measurement model (\ref{measvec}), the range domain is discretized into a number of range bins $\mathcal{R}=\left\{\tilde{r}_q\right\}_{q=1}^Q$ for subsequent processing, yielding an approximate linear measurement model.
Based on the range bin set $\mathcal{R}$, an overcomplete dictionary $\mathbf{A}(\tilde{\mathbf{r}})\in \mathbb{C}^{N\times Q}$ is constructed, and $\mathbf{A}(\tilde{\mathbf{r}}) = \left[\mathbf{s}_N(\tilde r_1), \mathbf{s}_N(\tilde r_2),\cdots,\mathbf{s}_N(\tilde r_Q) \right]$. Accordingly, when the targets lie exactly on the grid, the measurement model is
\begin{align}\label{measgrid}
        \mathbf{y} = \mathbf{A}(\tilde{\mathbf{r}}) \tilde{\mathbf{x}} + \mathbf{i} + \mathbf{w},
\end{align}
where $\tilde{\mathbf{x}} \in \mathbb{C}^{Q \times 1}$ denotes the sparse vector of target amplitudes, with $\|\tilde{\mathbf{x}}\|_0 = K$, where $\|\tilde{\mathbf{x}}\|_0$ denotes the number of nonzero elements of $\tilde{\mathbf{x}}$.
When the ranges of the targets are not exactly on the discretized grid, if the dictionary is sufficiently dense, model (\ref{measgrid}) is still valid and $\tilde{\mathbf{x}}$ is approximately sparse \cite{AWillsky}.




Let's discuss the model (\ref{measgrid}) briefly. If the jamming term $\mathbf{i}$ lacks inherent structure, the problem becomes ill-posed because, for any given $K$-sparse solution $\tilde{\mathbf{x}}$, a corresponding $\mathbf{i}$ can always be constructed. To enable target detection, it is necessary to impose a constraint on the jamming $\mathbf{i}$, namely that $\mathbf{i}$ is structured. For some challenging problems, it is often infeasible to develop tractable analytical models using handcrafted approaches. Nevertheless, with access to sufficient data, the underlying structure of the jamming can be learned, which in turn facilitates accurate jamming estimation as well as reliable target detection. In Section \ref{pretrain}, we present a methodology that implicitly learns the jamming distribution and explicitly estimates its score function from collected jamming data. Building upon this prior, we propose a model and data dual-driven framework in Section \ref{method}, which integrates the analytical echo model with the learned jamming score to achieve robust target detection under structured jamming.


\section{Pre-trained Neural Network for Jamming}\label{pretrain}
The structure of the jamming is often too complex to be adequately captured by analytical models. To facilitate its characterization, target-free jamming data are required for learning the underlying structure. In practice, such data can typically be acquired in two ways: either by generating jamming using simulation software, or by operating the radar in environments devoid of targets but subject to jamming.




The gradient of the log-likelihood $\log p(\mathbf{i})$ of $\mathbf{i}$ with respect to the conjugate of $\mathbf{i}$, i.e., $\nabla_{\mathbf{i}^{*}} \log p(\mathbf{i})$, is called the score function.
To capture the intrinsic structure of jamming signals, we pre-train a neural network within the diffusion model framework to learn the score function of $\mathbf{i}$. The network takes the jamming within a single pulse as input, which consists of $N$ fast time samples.
Aligning with common practices, network parameters are shared across all time steps. To accommodate standard convolutional neural networks, the complex signal is separated into its real and imaginary parts. Specifically, the input is organized as a two separate channel real valued tensor of size $N\times 2$, where the first channel is $\Re\{\mathbf{i}\}$ and the second channel is $\Im\{\mathbf{i}\}$. The neural network is designed to output a vector of dimension $N\times 2$. The first $N\times 1$ entries estimate the real component of the score, while the remaining $N\times 1$ estimate the imaginary component of the score. These outputs are then recombined into the complex valued score estimator
\begin{align}
    \mathbf{s}_{\hat\theta}(\mathbf{i}) = \Re(\mathbf{s}_{\hat\theta}(\mathbf{i})) + j\Im(\mathbf{s}_{\hat\theta}(\mathbf{i})) \in \mathbb{C}^{N\times1}.
\end{align}
Consequently, the two channel real value output of the network is naturally mapped to the complex valued score estimator $\mathbf{s}_{\hat\theta}(\mathbf{i})$, which is an approximation of $\nabla_{\mathbf{i}^{*}} \log p(\mathbf{i})$.

The diffusion model introduces a stochastic forward process that progressively perturbs the clean jamming signal $\mathbf{i}_0$, which corresponds to the jamming term $\mathbf{i}$ in measurement model (\ref{measgrid}). 
This evolution can be described by a stochastic differential equation (SDE) \cite{SongICLR} shown as
\begin{align}
\mathrm{d} \mathbf{i}_t= f(t)\mathbf{i}_t \mathrm{d} t+ g(t) \mathrm{d} \mathbf{w},
\end{align}
where $\mathbf{i}_t$ is the noised version data at time step $t$, $f(t)$ and $g(t)$ denote the drift and diffusion coefficients, respectively, $\mathbf{w}$ is a standard Wiener process, and $t\in[0,1]$.

Specifically, a Markov chain $\{\mathbf{i}_t\}_{t=0}^1$, where $t$ is randomly sampled from a uniform distribution over the interval $[0,1]$ with step size $\mathrm{d}t$, is constructed such that each step adds independent Gaussian noise according to
\begin{align}
    p(\mathbf{i}_t \mid \mathbf{i}_{t-\text{d} t}) = \mathcal{CN}\!\left(\mathbf{i}_t; \sqrt{1-\alpha_t}\,\mathbf{i}_{t-\text{d} t}, 2\alpha_t \mathbf{E}_{N}\right),
\end{align}
where $0<\alpha_t<1$ is a small noise increment coefficient,  $\mathbf{i}_t \mid \mathbf{i}_{t-\text{d} t}$ represents the perturbed jamming at time $t$ conditioned on the sample $\mathbf{i}_{t-\text{d} t}$, and $\mathbf{E}_N$ denotes the $N\times N$ identity matrix. Equivalently, one can write the sampling rule as \cite{Ho,JSDICML}
\begin{align}
    \mathbf{i}_t=\sqrt{1-\alpha_t} \mathbf{i}_{t-\text{d} t}+\sqrt{\alpha_t} \boldsymbol{\epsilon}_t, \quad \boldsymbol{\epsilon}_t \sim \mathcal{C N}(\mathbf{0}, 2\mathbf{E}_{N}).
\end{align}
After sufficient steps, the terminal state $\mathbf{i}_1$ converges to a complex Gaussian distribution, with its real and imaginary components follow standard Gaussian distribution, i.e., $q(\mathbf{i}_1)\simeq\mathcal{CN}(0,2\mathbf{E}_{N})$. Here, the symbol $\simeq$ denotes asymptotic equality \cite{asym}. Moreover, by marginalizing across the chain, one obtains the transition kernel \cite{Ho}
\begin{align}\label{kernelI}
p\left(\mathbf{i}_t \mid \mathbf{i}_0\right)=\mathcal{C N}\left(\mathbf{i}_t; \beta_t \mathbf{i}_0, 2b^2_t \mathbf{E}_{N}\right),
\end{align}
where ${\beta}_t = \prod_{s=1}^t \sqrt{1-\alpha_s}$ and $b^2_t=1-\beta^2_t$. This expression makes the relationship between any intermediate $\mathbf{i}_t$ and the original clean jamming $\mathbf{i}_0$ explicit.

\begin{figure}
	\centering
	\includegraphics[width = 5in]{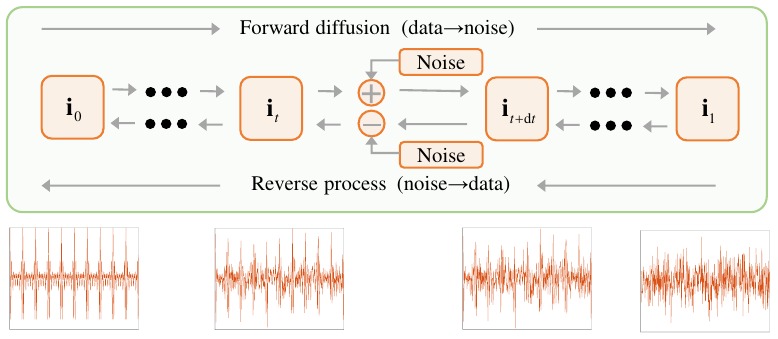}
	\caption{Diffusion process for jamming: In the forward process noise is gradually added to the clean data $\mathbf{i}_0$, while the reverse process learns to generate clean data starting from the corrupted data $\mathbf{i}_1$.}\label{DMprocess}
\end{figure}

The reverse diffusion process inverts this stochastic chain by iteratively denoising $\mathbf{i}_1$ back to $\mathbf{i}_0$, and a schematic illustration of the diffusion process is provided in Fig. \ref{DMprocess}. 
In practice, the reverse transitions are parameterized by a neural network $\mathbf{s}_\theta(\mathbf{i}_t,t)$ trained to estimate the score function $\nabla_{\mathbf{i}_t^*}\log p(\mathbf{i}_t)$ using denoising score matching (DSM), and the optimization problem is \cite{SongICLR}
\begin{align}\label{Iscorematch}
\hat{\boldsymbol{\theta}}=\underset{\boldsymbol{\theta}}{\operatorname{argmin}}\ \mathrm{E}_t\left\{\mathrm{E}_{\mathbf{i}_0} \mathrm{E}_{\mathbf{i}_t \mid \mathbf{i}_0}\left[\left\|\mathbf{s}_{\boldsymbol{\theta}}\left(\mathbf{i}_t, t\right)-\nabla_{\mathbf{i}_t^{*}} \log p\left(\mathbf{i}_t \mid \mathbf{i}_0\right)\right\|_2^2\right]\right\}.
\end{align}
This nested diffusion process enables the network converge to  the score function of the jamming, i.e., $\mathbf{s}_{\hat{\boldsymbol{\theta}}}\left(\mathbf{i}_t, t\right)\simeq \nabla_{\mathbf{i}_t^{*}} \log p\left(\mathbf{i}_t\right)$.

\begin{figure}
	\centering
	\includegraphics[width = 6in]{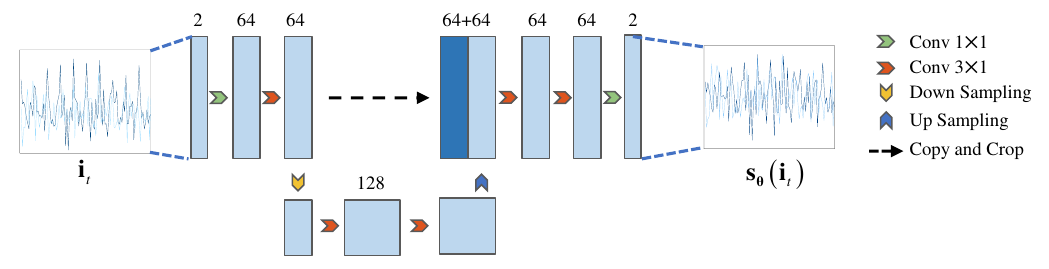}
	\caption{Structure of the DM’s denoising network using a lightweight U-net architecture.}\label{Unet}
\end{figure}

The U-Net architecture is capable of capturing waveform structures in the beam domain, as evidenced in \cite{unet}. 
Consequently, the denoising network is implemented as a lightweight U-Net \cite{unet}, which is shown in Fig. \ref{Unet}. The U-Net backbone is constructed with an encoder-decoder structure and skip connection to preserve fine scale features while enabling deep hierarchical representation learning. The encoder  downsamples the input data sequences through convolutional blocks, while the decoder performs symmetric upsampling with feature fusion. The network parameters are chosen to balance denoising performance and computational efficiency. Both the encoding and decoding process consist of 32 identical blocks of convolutional layer, each with a kernel size of $3$, group normalization, and rectified linear unit (ReLU) activation. Downsampling is performed using stride $2$ convolutions, with feature channels expanding from $64$ to $128$. The bottleneck layer maintains $128$ channels to extract global temporal features. The decoder symmetrically mirrors the encoder, employing transposed convolutions for upsampling and skip connection to retain high-resolution information, with channels contracting from $128$ back to $64$. Finally, a $1 \times 1$ convolution projects the features to the original input dimension, yielding the estimated jamming score.

Through this pretraining stage, the network learns to remove noise at each step, thereby enabling sampling of realistic jamming signals. From a probabilistic perspective, this corresponds to unconditional sampling from the learned prior distribution of jamming. Once the measurement model is incorporated, the procedure naturally extends to conditional sampling. Thus, in the next section, we aim to integrate the pre-trained jamming model $\mathbf{s}_{\hat{\boldsymbol{\theta}}}\left(\mathbf{i}_t, t\right)$ with the analytical measurement model (\ref{measgrid}) to estimate jamming and detect targets.

\section{Diffusion Based Model and Data Dual-driven Approach}\label{method}
\begin{figure}
	\centering
	\includegraphics[width = 6in]{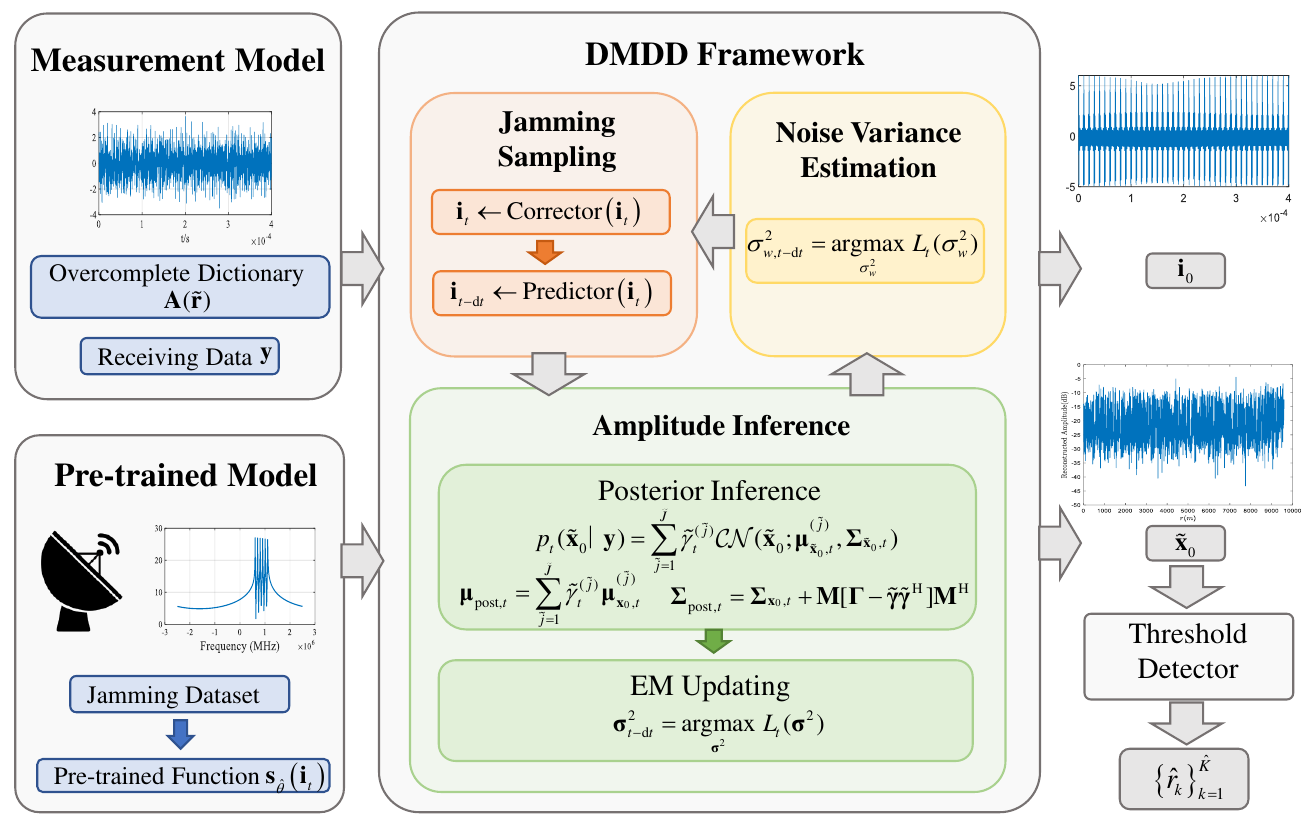}
	\caption{The proposed DMDD framework for handling multitargets detection under mainlobe jamming.}\label{flow}
\end{figure}
The targets are usually sparse in the range domain. To exploit the sparsity structure, the SBL is adopted. Specifically, the amplitude vector $\tilde{\mathbf{x}}_0$  is modeled as a zero-mean complex Gaussian vector with unknown variance  $\boldsymbol{\sigma}^2$ to enforce sparsity, i.e.,
\begin{align}\label{xprior}
    p(\tilde{\mathbf{x}}_0; \boldsymbol{\sigma}^2)=\mathcal{C} \mathcal{N}(\tilde{\mathbf{x}}_0 ; \mathbf{0}, \operatorname{diag}(\boldsymbol{\sigma}^2)).
\end{align}
The variable $\tilde{\mathbf{x}}_0$ corresponds to the clean variable at $t=0$, which is also identical to the amplitude vector $\tilde{\mathbf{x}}$ (\ref{measgrid}).

Based on the pre-trained jamming model $\mathbf{s}_{\hat{\boldsymbol{\theta}}}\left(\mathbf{i}_t, t\right)$ (\ref{Iscorematch}), and the linear measurement model (\ref{measgrid}), we propose a  model and data dual-driven approach named DMDD to integrate these components, including the sparsity prior $p(\tilde{\mathbf{x}}_0; \boldsymbol{\sigma}^2)$ (\ref{xprior}), where $\boldsymbol{\sigma}^2$ is unknown. In DMDD, an iterative mechanism is developed, which alternates between the reverse diffusion process for structured jamming estimation, the Bayesian inference of sparse target amplitudes and the noise variance estimation. The overall framework for multitarget detection under mainlobe jamming scenarios is illustrated in Fig. \ref{flow}, and the details of each module will be elaborated in the following subsections.


\subsection{Posterior Sampling of Jamming}


The end goal is to obtain samples from the posterior distribution $p(\mathbf{i}_0 \mid \mathbf{y})$, in which the intermediate steps are to sample from  the posterior distribution $p(\mathbf{i}_t \mid \mathbf{y})$ characterizing the structured jamming conditioned on the observed radar measurements. To this end, we formulate a conditional diffusion process $\{\mathbf{i}_t \mid \mathbf{y}\}_{t\in[0,1]}$, whose reverse-time dynamics are governed by the SDE given as
\begin{align}\label{reverse}
\mathrm{d}\left( \mathbf{i}_t\right)=\left[f(t) \mathbf{i}_t-g^2(t) \nabla_{\mathbf{i}^*_t} \log p\left(\mathbf{i}_t \mid \mathbf{y}\right)\right] \mathrm{d} t+g(t) \mathrm{d} \overline{\mathbf{w}}_t,
\end{align}
where $p\left(\mathbf{i}_t \mid \mathbf{y}\right)$ denotes the  posterior distribution of the jamming.
Let $\mathbf{i}_t^{(j)}$ denote the $j$th sample of  $\mathbf{i}_t$ conditioned on the measurements $\mathbf y$, $j=1,2,\cdots,J$. The final estimated jamming samples  $\mathbf{i}_0^{(j)}$ sampling from $p(\mathbf i_0|\mathbf y)$ can be created according to the reverse processes (\ref{reverse}). However, $\nabla_{\mathbf{i}^*_t} \log p\left(\mathbf{i}_t \mid \mathbf{y}\right)$ is unknown. 

According to the Bayesian rule, the posterior score can be decomposed as 
\begin{align}\label{Idiffusion}
\nabla_{\mathbf{i}_t^{(j) *}} \log p\left({\mathbf{i}}_t^{(j)} \mid \mathbf{y}\right) \simeq \nabla_{\mathbf{i}_t^{(j) *}} \log p\left(\mathbf{i}_t^{(j)}\right)+ \nabla_{\mathbf{i}_t^{(j) *}} \log p\left(\mathbf{y} \mid \mathbf{i}_t^{(j)}\right).
\end{align}
Here, the prior score term $\nabla_{\mathbf{i}_t^{(j)*}}\log p\left(\mathbf{i}_t^{(j)}\right)$ is approximated by the pretrained score network $\mathbf{s}_{\hat{\theta}}\left(\mathbf{i}_t^{(j)},t\right)$, while the second term requires accurate modeling of the noise-perturbed likelihood $p\left(\mathbf{y}\mid\mathbf{i}_t^{(j)}\right)$.
According to the measurement model (\ref{measgrid}), one can obtain the conditional distribution $p(\mathbf{y}\mid\mathbf{i}_0, \tilde{\mathbf{x}}_0)$, and at time step $t$, the variance parameter $\boldsymbol{\sigma}^2$ of the prior $p(\tilde{\mathbf{x}}_0; {\boldsymbol{\sigma}^2})$ is updated as $\boldsymbol{\sigma}^2_{t}$ by EM algorithm, which will be derived in Subsection \ref{PSx}.
By marginalizing $p(\mathbf{y}, \tilde{\mathbf{x}}_0|\mathbf{i}_0)=p(\mathbf{y}\mid\mathbf{i}_0, \tilde{\mathbf{x}}_0)p(\tilde{\mathbf{x}}_0; \boldsymbol{\sigma}^2_{t})$ over $\tilde{\mathbf{x}}_0$, one can derive
\begin{align}
p(\mathbf{y}\mid\mathbf{i}_0) = \int p(\mathbf{y}\mid\mathbf{i}_0, \tilde{\mathbf{x}}_0)p(\tilde{\mathbf{x}}_0; \boldsymbol{\sigma}^2_{t})\mathrm{d}\tilde{\mathbf{x}}_0.
\end{align}
Given the diffusion transition kernel $p(\mathbf{i}_t\mid\mathbf{i}_0)$ (\ref{kernelI}), the likelihood $p(\mathbf{y}\mid\mathbf{i}_t)$ at time step $t$ can be obtained by integrating over $\mathbf{i}_0$, i.e.,
\begin{align}
p(\mathbf{y}\mid\mathbf{i}_t) = \int p(\mathbf{y}\mid\mathbf{i}_0)p(\mathbf{i}_0\mid\mathbf{i}_t)\mathrm{d}\mathbf{i}_0.
\end{align}
However, computing $p(\mathbf{i}_0|\mathbf{i}_t) = \frac{p(\mathbf{i}_t|\mathbf{i}_0)p(\mathbf{i}_0)}{\int p(\mathbf{i}_t|\mathbf{i}_0)p(\mathbf{i}_0)d\mathbf{i}_0}$ requires knowledge of the intractable prior $p(\mathbf{i}_0)$, making direct calculation infeasible.


To overcome this issue, the Diffusion Model-based Posterior Sampling (DMPS) approach \cite{Meng2024} is adopted.
Based on the perturbation kernel (\ref{kernelI}), $\mathbf{i}_0$ can be expressed as
\begin{align}\label{IDMPS}
\mathbf{i}_0=\frac{\mathbf{i}_t-b_t \mathbf{w}_\mathbf{i}}{\beta_t},
\end{align}  
where $\mathbf{w}_\mathbf{i}\sim\mathcal{CN}(0,2\mathbf{E}_N)$ is AWGN. Substituting (\ref{IDMPS}) into the measurement model (\ref{measgrid}), an alternative representation of $\mathbf{y}$ can be expressed as
\begin{align}\label{appmodel}
\mathbf{y}=\mathbf{A} \tilde{\mathbf{x}}_0+\frac{1}{\beta_t} \mathbf{i}_t-\frac{b_t}{\beta_t} \mathbf{w}_\mathbf{i}+\mathbf{w}.
\end{align}
Define the equivalent noise $\mathbf{w}_{\text{eq},t}$ as
\begin{equation}\label{eqnoise}
\mathbf{w}_{\text{eq},t}=-\frac{b_t}{\beta_t} \mathbf{w}_\mathbf{i}+\mathbf{w},
\end{equation}
and model (\ref{appmodel}) reduces to 
\begin{align}\label{modelweq}
    \mathbf{y}=\mathbf{A} \tilde{\mathbf{x}}_0+\frac{1}{\beta_t} \mathbf{i}_t+\mathbf{w}_{\text{eq},t}.
\end{align}
Because the two noise terms, $\mathbf{w_i}$ and $\mathbf{w}$, are independent, with covariance matrix being $2\mathbf{E}_N$ and $\sigma_{w}^2\mathbf{E}_N$, respectively, the equivalent noise follows $\mathbf{w}_{\text{eq},t}\sim {\mathcal {CN}}({\mathbf 0},\sigma_{e,t}^2\mathbf{E}_{N})$ with 
\begin{align}\label{defsigmae}
\sigma_{e,t}^2=\frac{2b^2_t}{\beta^2_t}+\sigma_{w}^2.
\end{align}
And the conditional likelihood $p(\mathbf{y}\mid\mathbf{i}_t,\tilde{\mathbf{x}}_0)$ follows
\begin{align}\label{eq:cond_y_i_x}
p(\mathbf{y}\mid\mathbf{i}_t,\tilde{\mathbf{x}}_0)
= \mathcal{CN}\left(\mathbf{y};\mathbf{A}\tilde{\mathbf{x}}_0 + \tfrac{1}{\beta_t}\mathbf{i}_t,\sigma_{e,t}^2\mathbf{E}_N\right).
\end{align}

At time step $t$, the prior distribution of $\tilde{\mathbf{x}}_0$ (\ref{xprior}) is employed, whose variances $\boldsymbol{\sigma}^2$ are updated as $\boldsymbol{\sigma}^2_t$, and the details is deferred to Subsection \ref{PSx}. Based on this prior, the likelihood $p\left(\mathbf{y}\mid\mathbf{i}_t\right)$ is obtained by marginalizing $p(\mathbf{y},\tilde{\mathbf{x}}_0\mid\mathbf{i}_t)=p(\mathbf{y}\mid\mathbf{i}_t,\tilde{\mathbf{x}}_0)
p(\tilde{\mathbf{x}}_0; \boldsymbol{\sigma}^2_{t})$ over $\tilde{\mathbf{x}}_0$, i.e.,
\begin{align}\label{eq:marginal_y_i}
p(\mathbf{y}\mid\mathbf{i}_t)
= \int p(\mathbf{y}\mid\mathbf{i}_t,\tilde{\mathbf{x}}_0)
p(\tilde{\mathbf{x}}_0; \boldsymbol{\sigma}^2_{t})\mathrm{d}\tilde{\mathbf{x}}_0.
\end{align}
Since both distributions $p(\mathbf{y}\mid\mathbf{i}_t,\tilde{\mathbf{x}}_0)$  (\ref{eq:cond_y_i_x}) and $p(\tilde{\mathbf{x}}_0; \boldsymbol{\sigma}^2_{t})$ (\ref{xprior}) are complex Gaussian, the marginalization $p(\mathbf{y}\mid\mathbf{i}_t)$ (\ref{eq:marginal_y_i}) admits a closed-form expression, resulting in another complex Gaussian distribution
\begin{align}\label{eq:final_py_it}
p(\mathbf{y}\mid\mathbf{i}_t)
= \mathcal{CN}\left(
\mathbf{y};
\boldsymbol{\mu}_{\mathbf{y},t},
\boldsymbol{\Sigma}_{\mathbf{y},t}
\right),
\end{align}
where the mean and covariance are given by $\boldsymbol{\mu}_{\mathbf{y},t} = \frac{1}{\beta_t} \mathbf{i}_t$, and $\boldsymbol{\Sigma}_{\mathbf{y},t} = \mathbf{A} \operatorname{diag}(\boldsymbol{\sigma}^2_{t})
 \mathbf{A}^{\mathrm{H}} + \sigma_{e,t}^2\mathbf{E}_{N}$.


The corresponding closed-form likelihood score functions are then obtained as
\begin{align}\label{scoreresult}
\nabla_{\mathbf{i}_t^{*}} \log p\left(\mathbf{y} \mid \mathbf{i}_t\right)=\frac{1}{\beta_t} \boldsymbol{\Sigma}_{\mathbf{y}, t}^{-1}\left(\mathbf{y}-\boldsymbol{\mu}_{\mathbf{y},t}\right).
\end{align}
Finally, substituting (\ref{scoreresult}) into the posterior score decomposition (\ref{Idiffusion}), the updated posterior scores can be expressed as
\begin{align}\label{samplei}
\nabla_{\mathbf{i}_t^{(j) *}} \log p\left(\mathbf{i}_t^{(j)} \mid \mathbf{y}\right) \simeq \mathbf{s}_{\hat{\theta}}\left(\mathbf{i}_t^{(j)}, t\right) +\frac{1}{\beta_t}  \boldsymbol{\Sigma}_{\mathbf{y}, t}^{-1}\left(\mathbf{y}-\frac{1}{\beta_t} \mathbf{i}_t^{(j)}\right).
\end{align}
And the predictor and corrector sampling algorithm is implemented \cite{SongICLR}. At each iteration, the corrector performs Langevin dynamics to refine the jamming samples $\mathbf{i}_t$, and the update rule is
\begin{align}
    \mathbf{i}_t^{(j)} \leftarrow \mathbf{i}_t^{(j)} + \epsilon_t \nabla_{\mathbf{i}_t^{(j) *}} \log p\left(\mathbf{i}_t^{(j)} \mid \mathbf{y}\right) + \sqrt{2 \epsilon_t} \mathbf{z}_{\mathbf{i}},
\end{align}
where $\epsilon_t$ is the step size, and $\mathbf{z}_{\mathbf{i}} \sim \mathcal{C N}\left(\mathbf{0}, 2 \mathbf{E}_{N}\right)$.
The predictor step then follows the reverse-time SDE to propagate the samples, and the update rule is
\begin{align}
    \mathbf{i}_{t-\text{d}t}^{(j)} \leftarrow \mathbf{i}_{t}^{(j)} +\left(-f(t) \mathbf{i}_t^{(j)} + g^2(t) \nabla_{\mathbf{i}_t^{(j) *}} \log p\left(\mathbf{i}^{(j)}_t \mid \mathbf{y}\right)\right)\text{d}t + g(t)\sqrt{ \text{d}t} \mathbf{z}_{\mathbf{i}}.
\end{align}
For further details about the implementation of the predictor and corrector sampler, please see \cite{SongICLR}.

Note that $\nabla_{\mathbf{i}_t^{(j) *}} \log p\left(\mathbf{i}_t^{(j)} \mid \mathbf{y}\right)$ (\ref{scoreresult}) involves the unknown parameters ${\boldsymbol\sigma}_t^2$ and $\sigma_{e,t}^2$, which have to be estimated during the iteration. Details are given in the ensuing subsections.
\subsection{Posterior Inference of Amplitudes}\label{PSx}
In this subsection, once the jamming samples $\{\mathbf{i}_t^{(j)}\}_{j=1}^J$ have been sampled at time step $t$,  posterior estimation of the target amplitudes is subsequently carried out within a model-based framework. This estimated posterior will then be used to update the hyperparameter $\boldsymbol{\sigma}^2$ of the target amplitude prior (\ref{xprior}).



Based on model (\ref{modelweq}), and given the estimated jamming term at each iteration, the conditional likelihood function can be written as
\begin{align}
p_t(\mathbf{y}\mid\tilde{\mathbf{x}}_0,\mathbf{i}_t)=\mathcal{CN}(\mathbf{y},\mathbf{A} \tilde{\mathbf{x}}_0+\frac{1}{\beta_t} \mathbf{i}_t,\sigma_{e,t}^2 \mathbf{E}_{N}).
\end{align}
By marginalizing $p_t(\mathbf{y},\tilde{\mathbf{x}}_0\mid\mathbf{i}_t)=p_t(\mathbf{y}\mid\tilde{\mathbf{x}}_0,\mathbf{i}_t)p(\mathbf{i}_t)$ over $\mathbf{i}_t$, the likelihood $p_t(\mathbf{y}\mid\tilde{\mathbf{x}}_0)$ is obtained as
\begin{align}\label{py_x0}
    p_t(\mathbf{y}\mid\tilde{\mathbf{x}}_0)=\int p_t(\mathbf{y}\mid\tilde{\mathbf{x}}_0,\mathbf{i}_t)p(\mathbf{i}_t)\text{d} \mathbf{i}_t.
\end{align}
$p(\mathbf{i}_t)$ is provided by the neural network and does not admit a closed-form expression. The $\tilde{J}$ independent samples $\tilde{\mathbf{i}}_t^{(\tilde{j})}$ are generated according to the original data $\mathbf{i}_0$ and the transition $p(\mathbf{i}_t|\mathbf{i}_0)$ (\ref{kernelI}).
Therefore, we approximate $p(\mathbf{i}_t)$ as
\begin{align}\label{appit}
    p(\mathbf{i}_t) \approx \frac{1}{\tilde{J}}\sum_{j=1}^{\tilde{J}} \delta(\mathbf{i}_t - \tilde{\mathbf{i}}_t^{(\tilde{j})}).
\end{align}
Substituting (\ref{appit}) into (\ref{py_x0}) yields
\begin{align}
\begin{aligned}
p_t(\mathbf{y}\mid\tilde{\mathbf{x}}_0)
&\approx \int p_t(\mathbf{y}\mid\tilde{\mathbf{x}}_0,\mathbf{i}_t)\left(\frac{1}{\tilde{J}}\sum_{\tilde{j}=1}^{\tilde{J}}\delta(\mathbf{i}_t-\tilde{\mathbf{i}}_t^{(\tilde{j})})\right)\mathrm{d}\mathbf{i}_t \\
&= \frac{1}{\tilde{J}}\sum_{\tilde{j}=1}^{\tilde{J}} \int p_t(\mathbf{y}\mid\tilde{\mathbf{x}}_0,\mathbf{i}_t)\delta(\mathbf{i}_t-\tilde{\mathbf{i}}_t^{(\tilde{j})})\mathrm{d}\mathbf{i}_t \\
&= \frac{1}{\tilde{J}}\sum_{\tilde{j}=1}^{\tilde{J}} p_t\big(\mathbf{y}\mid\tilde{\mathbf{x}}_0,\tilde{\mathbf{i}}_t^{(\tilde{j})}\big),
\end{aligned}
\end{align}
which is a Gaussian mixture model (GMM) and can be rewritten as
\begin{align}\label{GMM}
    p_t(\mathbf{y}\mid\tilde{\mathbf{x}}_0)\approx \frac{1}{\tilde{J}}\sum_{\tilde{j}=1}^{\tilde{J}}\mathcal{CN}(\mathbf{y};\tilde{\boldsymbol{\mu}}_{\mathbf{y},t}^{(\tilde{j})},\sigma_{e,t}^2 \mathbf{E}_{N})\triangleq p_t^{\text{app}}(\mathbf{y}\mid\tilde{\mathbf{x}}_0),
\end{align}
where the mean of the $\tilde{j}$th component in $p_t^{\text{app}}(\mathbf{y}\mid\tilde{\mathbf{x}}_0)$ is $\tilde{\boldsymbol{\mu}}_{\mathbf{y},t}^{(\tilde{j})}=\mathbf{A} \tilde{\mathbf{x}}_0+\frac{1}{\beta_t} \tilde{\mathbf{i}}^{(\tilde{j})}_t$.
This representation shows that the marginalized likelihood can be expressed as a Gaussian mixture with a common covariance matrix and distinct component means.

According to the Bayesian rule, at time step $t$, the posterior distribution $p_t(\tilde{\mathbf{x}}_0 \mid \mathbf{y})$ is obtained as
\begin{align}
\begin{aligned}
    p_t(\tilde{\mathbf{x}}_0 \mid \mathbf{y}) &\propto p_t^{\text{app}}(\mathbf{y} \mid \tilde{\mathbf{x}}_0) p(\tilde{\mathbf{x}}_0;\boldsymbol{\sigma}_{t}^2)\\
    &=  \frac{1}{\tilde{J}}\sum_{\tilde{j}=1}^{\tilde{J}} \mathcal{CN}(\mathbf{y};\tilde{\boldsymbol{\mu}}_{\mathbf{y},t}^{(\tilde{j})},\sigma_{e,t}^2 \mathbf{E}_{N})\mathcal{C} \mathcal{N}(\tilde{\mathbf{x}}_0 ; \mathbf{0}, \operatorname{diag}(\boldsymbol{\sigma}_{t}^2)).
\end{aligned}
\end{align}
Since the product of Gaussian densities remains Gaussian, multiplying the Gaussian prior with the Gaussian mixture likelihood yields another Gaussian mixture. Hence,
\begin{align}
    p_t(\tilde{\mathbf{x}}_0 \mid \mathbf{y})\propto\sum_{\tilde{j}=1}^{\tilde{J}} \gamma_t^{(\tilde{j})} \mathcal{CN}(\tilde{\mathbf{x}}_0 ;\tilde{\boldsymbol{\mu}}_{\tilde{\mathbf{x}}_0,t}^{(\tilde{j})}, \tilde{\boldsymbol{\Sigma}}_{\tilde{\mathbf{x}}_0,t}),
\end{align}
where the means and covariances of the $j$th component in $p_t(\tilde{\mathbf{x}}_0 \mid \mathbf{y})$ are given by
\begin{align}
&\tilde{\boldsymbol{\mu}}_{\tilde{\mathbf{x}}_0,t}^{(\tilde{j})}=\tilde{\boldsymbol{\Sigma}}_{\tilde{\mathbf{x}}_0,t}\boldsymbol{\phi}_t^{(\tilde{j})},\\
&\tilde{\boldsymbol{\Sigma}}_{\tilde{\mathbf{x}}_0,t}= \left(\frac{1}{\sigma_{e,t}^2} \mathbf{A}^{\rm H} \mathbf{A} +\operatorname{diag}^{-1}(\boldsymbol{\sigma}_{t}^2)\right)^{-1},
\end{align}
where $\boldsymbol{\phi}_t^{(\tilde{j})} = \frac{1}{\sigma_{e,t}^2}  \mathbf{A}^\mathrm{H} \left( \mathbf{y} - \frac{1}{\beta_t} \tilde{\mathbf{i}}_t^{(\tilde{j})} \right)$.
The weight coefficient $\gamma_t^{(\tilde{j})}$ is given by
\begin{align}
    \gamma_t^{(\tilde{j})} = \frac{1}{\tilde{J}} \frac{\operatorname{det}(\tilde{\boldsymbol{\Sigma}}_{\tilde{\mathbf{x}}_0,t})}{\pi^{N}\operatorname{det}\left(\sigma_{e,t}^2 \mathbf{E}_{N}\right) \operatorname{det}\left(\operatorname{diag}(\boldsymbol{\sigma}_{t}^2)\right)} e^{-\kappa^{(\tilde{j})}_t},
\end{align}
where $\kappa^{(\tilde{j})}_t=\psi^{(\tilde{j})}_t-\boldsymbol{\phi}^{(\tilde{j})\rm H}_t \tilde{\boldsymbol{\Sigma}}_{\tilde{\mathbf{x}}_0,t} \boldsymbol{\phi}^{(\tilde{j})}_t$ and $\psi^{(\tilde{j})}_t=\sigma_{e,t}^2\left( \mathbf{y} - \frac{1}{\beta_t} \tilde{\mathbf{i}}_t^{(\tilde{j})} \right)^{\rm H}  \left( \mathbf{y} - \frac{1}{\beta_t} \tilde{\mathbf{i}}_t^{(\tilde{j})} \right)$.

By defining the normalized weight coefficients $\tilde{\gamma}_t^{(\tilde{j})}$ as 
\begin{align}
\begin{aligned}
\tilde{\gamma}_t^{(\tilde{j})}=&\gamma_t^{(\tilde{j})}/\sum_{\tilde{j}=1}^{\tilde{J}}\gamma_t^{(\tilde{j})}\\
=&\frac{e^{-\kappa^{(\tilde{j})}_t}}{\sum_{k=1}^{\tilde{J}} e^{-\kappa^{(k)}_t}}\\
=&\frac{1}{1+\sum_{k \neq \tilde{j}} e^{\kappa^{(\tilde{j})}_t-\kappa^{(k)}_t}}, 
\end{aligned}
\end{align} 
the posterior distribution $ p_t(\tilde{\mathbf{x}}_0 \mid \mathbf{y})$ is
\begin{align}
    p_t(\tilde{\mathbf{x}}_0 \mid \mathbf{y})=\sum_{\tilde{j}=1}^{\tilde{J}} \tilde{\gamma}_t^{(\tilde{j})} \mathcal{CN}(\tilde{\mathbf{x}}_0 ;\tilde{\boldsymbol{\mu}}_{\tilde{\mathbf{x}}_0,t}^{(\tilde{j})}, \tilde{\boldsymbol{\Sigma}}_{\tilde{\mathbf{x}}_0,t}).
\end{align}
The posterior mean and posterior covariance matrix of $\tilde{\mathbf{x}}_0$ are calculated as
\begin{align}
\begin{aligned}\label{mupost}
\boldsymbol{\mu}_{\text{post},t} &= \int\tilde{\mathbf{x}}_0 p_t(\tilde{\mathbf{x}}_0 \mid \mathbf{y}) \rm{d} \tilde{\mathbf{x}}_0\\
    &=\sum_{\tilde{j}=1}^{\tilde{J}} \tilde{\gamma}_t^{(\tilde{j})} \tilde{\boldsymbol{\mu}}_{\tilde{\mathbf{x}}_0,t}^{(\tilde{j})},
\end{aligned}
\end{align}
and 
\begin{align}\label{sigmapost}
    \begin{aligned}
        \boldsymbol{\Sigma}_{\text{post},t} &= \int \tilde{\mathbf{x}}_0\tilde{\mathbf{x}}_0^{\rm H} p(\tilde{\mathbf{x}}_0 \mid \mathbf{y})\rm{d} \tilde{\mathbf{x}}_0 - \boldsymbol{\mu}_{\text{post},t}\boldsymbol{\mu}_{\text{post},t}^{\rm H}\\
&=\tilde{\boldsymbol{\Sigma}}_{\tilde{\mathbf{x}}_0,t}+\sum_{\tilde{j}=1}^{\tilde{J}} \tilde{\gamma}_t^{(\tilde{j})}  \tilde{\boldsymbol{\mu}}_{\tilde{\mathbf{x}}_0,t}^{(\tilde{j})} \tilde{\boldsymbol{\mu}}_{\tilde{\mathbf{x}}_0,t}^{(\tilde{j})\rm H}-\boldsymbol{\mu}_{\text{post},t}\boldsymbol{\mu}_{\text{post},t}^{\rm H}\\
&=\tilde{\boldsymbol{\Sigma}}_{\tilde{\mathbf{x}}_0,t}+\mathbf{M}[\boldsymbol{\Gamma}-\tilde{\boldsymbol{\gamma}} \tilde{\boldsymbol{\gamma}}^{\rm H}]\mathbf{M}^{\rm H},
    \end{aligned}
\end{align}
where $\mathbf{M}=\left[\tilde{\boldsymbol{\mu}}_{\tilde{\mathbf{x}}_0,t}^{(1)},\tilde{\boldsymbol{\mu}}_{\tilde{\mathbf{x}}_0,t}^{(2)},\cdots,\tilde{\boldsymbol{\mu}}_{\tilde{\mathbf{x}}_0,t}^{(\tilde{J})}\right]$, $\boldsymbol{\Gamma}=\operatorname{diag}\left(\tilde{\gamma}_t^{(1)} ,\tilde{\gamma}_t^{(2)} ,\cdots,\tilde{\gamma}_t^{(\tilde{J})} \right)$, and $\tilde{\boldsymbol{\gamma}}=\left[\tilde{\gamma}_t^{(1)} ,\tilde{\gamma}_t^{(2)} ,\cdots,\tilde{\gamma}_t^{(\tilde{J})} \right]^{\rm T}$.

Furthermore, at time step $t$ in which the hyperparameter $\boldsymbol{\sigma}^2$ is estimated as $\boldsymbol{\sigma}^2_{t}$, hyperparameter $\boldsymbol{\sigma}^2$ is updated as $\boldsymbol{\sigma}^2_{t-{\rm d}t}$ via the EM algorithm. By maximizing the expected complete loglikelihood, where the expectation is taken with respect to the posterior of $\tilde{\mathbf{x}}_0$, $\boldsymbol{\sigma}^2_{t-{\rm d}t}$ is updated as 
\begin{align}
    {\boldsymbol{\sigma}}^2_{t-{\rm d}t}=\underset{\boldsymbol{\sigma}^2} {\operatorname{argmax}}\ 
     L_t(\boldsymbol{\sigma}^2),
\end{align}
where the objective function $L_t(\boldsymbol{\sigma}^2)$ is given by 
\begin{align}\label{expall}
\begin{aligned}
L_t(\boldsymbol{\sigma}^2)&\triangleq \mathrm{E}_{ \tilde{\mathbf{x}}_0\mid \mathbf{y}}\left[
\log p(\mathbf{y}, \tilde{\mathbf{x}}_0; \boldsymbol{\sigma}^2)\right]\\
&= \mathrm{E}_{ \tilde{\mathbf{x}}_0\mid \mathbf{y}}\left[\log p(\mathbf{y}\mid\tilde{\mathbf{x}}_0)
+ \log p(\tilde{\mathbf{x}}_0; \boldsymbol{\sigma}^2)\right].
\end{aligned}
\end{align} 
Since the likelihood term $p(\mathbf{y}\mid\tilde{\mathbf{x}}_0)$ is irrelevant with the hyperparameter $\boldsymbol{\sigma}^2$, it can be dropped when maximizing $L_t(\boldsymbol{\sigma}^2)$. Therefore, (\ref{expall}) can be simplified as
\begin{align}
\begin{aligned}
    L_t(\boldsymbol{\sigma}^2)&= \mathrm{E}_{ \tilde{\mathbf{x}}_0\mid \mathbf{y}}\left[\log p(\tilde{\mathbf{x}}_0 ; \boldsymbol{\sigma}^2)\right]+\text{const}_1= \mathrm{E}\left[\tilde{\mathbf{x}}_0^{\rm H}\operatorname{diag}^{-1}(\boldsymbol{\sigma}^2)\tilde{\mathbf{x}}_0\right] - \operatorname{log}|\operatorname{diag}(\boldsymbol{\sigma}^2)|+\text{const}_2\\
    &=-\sum_{k=1}^{Q} \frac{\mathrm{E}\left[\left|\tilde{\mathbf{x}}_0(k)\right|^2\right]}{\boldsymbol{\sigma}^2(k)}-\log \prod_{k=1}^{Q}\left( \boldsymbol{\sigma}^2(k)\right)+\text{const}_2 \\
    & =-\sum_{k=1}^{Q} \frac{\boldsymbol{\Sigma}_{\text{post},t}(k)+\left|\boldsymbol{\mu}_{\text{post},t}(k)\right|^2}{\boldsymbol{\sigma}^2(k)}-\sum_{k=1}^{Q} \log \left(\boldsymbol{\sigma}^2(k)\right)+\text{const}_2,
\end{aligned}
\end{align} 
where both $\text{const}_1$ and $\text{const}_2$ denote terms that are independent of $\boldsymbol{\sigma}^2$,  $\mathrm{E}_{ \tilde{\mathbf{x}}_0\mid \mathbf{y}}[\cdot]$ denotes the expectation with respect to the conditional probability density function $p_t(\tilde{\mathbf{x}}_0 \mid \mathbf{y})$.


Taking the derivative of $L_t(\boldsymbol{\sigma}^2)$ with respect to $\boldsymbol{\sigma}^2(k)$ yields
\begin{align}
    \frac{\partial L_t(\boldsymbol{\sigma}^2)}{\partial \boldsymbol{\sigma}^2(k)}=\frac{\boldsymbol{\Sigma}_{\text{post},t}(k)+\left|\boldsymbol{\mu}_{\text{post},t}(k)\right|^2}{\left(\boldsymbol{\sigma}^2(k)\right)^2}-\frac{1}{ \boldsymbol{\sigma}^2(k)} .
\end{align}
Setting this derivative to zero gives the closed-form update rule, and ${\boldsymbol{\sigma}}^2_{t-{\rm d}t}$ is updated as 
\begin{align}
    {\boldsymbol{\sigma}}^2_{t-{\rm d}t} = \boldsymbol{\Sigma}_{\text{post},t}+\left|\boldsymbol{\mu}_{\text{post},t}\right|^{\odot 2}.
\end{align}

\subsection{Estimation of Noise Variance}
The noise variance $\sigma_w^2$ is also estimated at time step $t$ using the EM algorithm. 
Based on the GMM (\ref{GMM}), the noise variance $\sigma_{w}^2$ can be updated as ${\sigma}^2_{w,t-{\rm d}t}$ at time step $t$ by maximizing the expected log-likelihood
shown as
\begin{align}
    {\sigma}^2_{w,t-{\rm d}t}=\underset{\sigma^2_{w}} {\operatorname{argmax}} \ L_t(\sigma^2_{w}),
\end{align}
where the objective function is given by
\begin{align}\label{Lt}
    \begin{aligned}
        L_t(\sigma^2_{w})&\triangleq \mathrm{E}_{ \tilde{\mathbf{x}}_0\mid \mathbf{y}}\left[\log p_t(\mathbf{y}\mid \tilde{\mathbf{x}}_0)\right]\\
        &\approx \mathrm{E}_{ \tilde{\mathbf{x}}_0\mid \mathbf{y}}\left[\log\left(\frac{1}{\tilde{J}}\sum_{\tilde{j}=1}^{\tilde{J}} \mathcal{CN}\left(\mathbf{y};\tilde{\boldsymbol{\mu}}_{\mathbf{y},t}^{(\tilde{j})},\left(\sigma_{w}^2+\frac{2b_t^2}{\beta_t^2}\right) \mathbf{E}_{N}\right)\right)\right]\triangleq L_t^{\rm app}(\sigma^2_{w}).
\end{aligned}\end{align}

It is difficult to derive a closed-form expression of $L_t^{\rm app}(\sigma^2_{w})$ (\ref{Lt}). As a practical alternative, a heuristic approach is adopted to obtain its lower bound instead.

For a convex function $f$, if $\mathbf{x}$ are random variables such that $\mathbf{x} \in \mathrm{dom}\ f$ with probability one, where $\mathrm{dom}\ f$ denotes the domain of $f$, then Jensen’s inequality states that
$f\big(\mathbb{E}[\mathbf{x}]\big) \leq \mathbb{E}[f(\mathbf{x})]$.
Since the logarithm function $\log(\cdot)$ is concave, the inequality is reversed, i.e.,
\begin{align}\label{jensen}
    \log\big(\mathbb{E}[\mathbf{x}]\big) \geq \mathbb{E}[\log(\mathbf{x})].
\end{align}
Applying the inequality (\ref{jensen}) to $L_t^{\rm app}(\sigma^2_{w})$ (\ref{Lt}), one has
\begin{align}\label{Jesen}
    \log\left(\frac{1}{\tilde{J}}\sum_{\tilde{j}=1}^{\tilde{J}} \mathcal{CN}\left(\mathbf{y};\tilde{\boldsymbol{\mu}}_{\mathbf{y},t}^{(\tilde{j})},\left(\sigma_{w}^2+\frac{2b_t^2}{\beta_t^2}\right) \mathbf{E}_{N}\right)\right) \geq \frac{1}{\tilde{J}}\sum_{\tilde{j}=1}^{\tilde{J}} \log\left(\mathcal{CN}\left(\mathbf{y};\tilde{\boldsymbol{\mu}}_{\mathbf{y},t}^{(\tilde{j})},\left(\sigma_{w}^2+\frac{2b_t^2}{\beta_t^2}\right) \mathbf{E}_{N}\right)\right).
\end{align}
Substituting (\ref{Jesen}) into (\ref{Lt}), a tractable lower bound $L_b(\sigma^2_{w})$ of $L_t^{\rm app}(\sigma^2_{w})$ can be established as
\begin{align}
    \begin{aligned}
        L_t^{\rm app}(\sigma^2_{w})&\geq L_b(\sigma^2_{w})\\
        &= \mathrm{E}_{ \tilde{\mathbf{x}}_0\mid \mathbf{y}}\left[\frac{1}{\tilde{J}}\sum_{\tilde{j}=1}^{\tilde{J}} \log\left(\mathcal{CN}\left(\mathbf{y};\tilde{\boldsymbol{\mu}}_{\mathbf{y},t}^{(\tilde{j})},\left(\sigma_{w}^2+\frac{2b_t^2}{\beta_t^2}\right) \mathbf{E}_{N}\right)\right)\right]\\
        &= \frac{1}{\tilde{J}}\sum_{\tilde{j}=1}^{\tilde{J}}-\frac{\mathrm{E}_{ \tilde{\mathbf{x}}_0\mid \mathbf{y}}\left[\|\mathbf{y}-\mathbf{A}\tilde{\mathbf{x}}_0-\tilde{\mathbf{i}}_t^{(\tilde{j})}\|^2_2\right]}{\sigma_{w}^2+\frac{2b_t^2}{\beta_t^2}} - N \log \left(\sigma_{w}^2+\frac{2b_t^2}{\beta_t^2}\right) + \text{const}\\
        &=\frac{1}{\tilde{J}}\sum_{\tilde{j}=1}^{\tilde{J}}-\frac{\|\mathbf{y}-\mathbf{A} \boldsymbol{\mu}_{\text{post},t} -\tilde{\mathbf{i}}_t^{(\tilde{j})}\|^2_2 + \operatorname{tr}(\mathbf{A} \boldsymbol{\Sigma}_{\text{post},t}\mathbf{A}^{\rm H})}{\sigma_{w}^2+\frac{2b_t^2}{\beta_t^2}} - N \log \left(\sigma_{w}^2+\frac{2b_t^2}{\beta_t^2}\right) + \text{const},
    \end{aligned}
\end{align}
where $\text{const}$ denotes terms that are independent of $\sigma^2_{w}$. We maximize the lower bound $L_b(\sigma^2_{w})$ instead by taking the derivative of $L_b(\sigma^2_{w})$ with respect to $\sigma^2_{w}$, yielding
\begin{align}
    \frac{\partial L_b(\sigma^2_{w})}{\partial \sigma^2_{w}}=\sum_{\tilde{j}=1}^{\tilde{J}}\frac{\|\mathbf{y}-\mathbf{A} \boldsymbol{\mu}_{\text{post},t} -\tilde{\mathbf{i}}_t^{(\tilde{j})}\|^2_2 + \operatorname{tr}(\mathbf{A} \boldsymbol{\Sigma}_{\text{post},t}\mathbf{A}^{\rm H})}{\tilde{J}\left(\sigma_{w}^2+\frac{2b_t^2}{\beta_t^2}\right)^2}-\frac{N}{\sigma_{w}^2+\frac{2b_t^2}{\beta_t^2}} .
\end{align}
Setting this derivative to zero and enforcing the nonnegativity of $\sigma_{w}^2$ give the estimation of $\sigma_{w,t-\text{d}t}^2$, which can be calculated as
\begin{align}
    {\sigma}^2_{w,t-\text{d}t}=\min\left(\frac{1}{N}\left(\frac{1}{\tilde{J}}\left(\sum_{\tilde{j}=1}^{\tilde{J}}\|\mathbf{y}-\mathbf{A} \boldsymbol{\mu}_{\text{post},t} -\tilde{\mathbf{i}}_t^{(\tilde{j})}\|^2_2\right) + \operatorname{tr}(\mathbf{A} \boldsymbol{\Sigma}_{\text{post},t}\mathbf{A}^{\rm H})\right)-\frac{2b_t^2}{\beta_t^2},\zeta\right)
    ,
\end{align}
where $\min(a,b)$ returns the minimum of $a$ and $b$, $\zeta$ is a small preset positive parameter.
The updated noise variance is then used for the next iteration. Numerical experiments verify that this heuristic approach yields satisfactory performance.

In this work, instead of employing an adaptive threshold detector, a constant threshold detector is adopted after obtaining the estimate of $\tilde{\mathbf{x}}_0$. That is, the detection threshold is independent of the estimated amplitude $\tilde{\mathbf{x}}_0$, and its value is determined based on the receiver sensitivity. Note that the proposed DMDD method estimates the noise variance $\sigma_w^2$ and the target amplitude $\tilde{\mathbf{x}}_0$. The target power relative to the noise level can be calculated as $P=10\log (\frac{|\tilde{\mathbf{x}}_0|^2}{\sigma_w^2})$. Assuming that the minimum detectable target power is $P_{\rm min}$, the detection threshold is set as $P_{\rm min} = (T_h - G_a)$ dB, where $G_a = 10\log N_p$ dB denotes the signal processing gain at the receiver with $N_p$ being the number of coherent samples contributing to the integration gain, and $T_h$ is a carefully defined parameter determined according to the practical system configuration.

In summary, the DMDD algorithm begins with a pretraining stage, where the score model of the jamming is learned from the dataset.
During each iteration of the sampling stage, the pretrained score model is utilized to refine the jamming samples through a corrector step, followed by a predictor update. Subsequently, the posterior inference of the target amplitude vector is carried out. The EM algorithm is then employed to update the prior hyperparameters of the target amplitudes and the noise variance. These steps are iteratively repeated until the diffusion time step reaches $t=0$, indicating the completion of the sampling stage. Finally, a constant threshold detector is applied.
And the overall algorithm is summarized in Algorithm \ref{algor}.

\begin{algorithm}
\renewcommand{\algorithmicrequire}{\textbf{Input:}}
\caption{DMDD}
\label{algor}
\begin{algorithmic}[1]
\REQUIRE $\mathbf{y}$, $\mathbf{A}$, $\mathbf{s}_{\hat\theta}\left(\mathbf{i}_t,t\right)$,  $T$, $J$, $\tilde{J}$, $\{\beta_t\}_{t=0}^1$, $f(t)$, $g(t)$, $\zeta$
    \STATE \textbf{Initialize:}  $\mathbf{i}_1^{(j)} \sim \mathcal{C N}(\mathbf{0},\left(2\left(1-\beta_1^2\right) \mathbf{E}_{N}\right)$, $j=1,2,\cdots, J$, $\boldsymbol{\sigma}^2_{1} \leftarrow 10^4\times \mathbf{1}_{Q}$, $\mathrm{~d} t \leftarrow \frac{1}{T}$, $\sigma_w^2\leftarrow1$
    \STATE  Generate $\tilde{J}$ samples $\{\tilde{\mathbf{i}}_t^{(\tilde{j})}\}_{t=0}^1$ according to (\ref{kernelI}) and the original data $\mathbf{i}_0$, where $\tilde{j}=1,2,\cdots, \tilde{J}$ 
    \FOR {$m=T-1 \text { to } 0$}
    \STATE $t \leftarrow \frac{m+1}{T}$
    \STATE Calculate $\nabla_{\mathbf{i}_t^{(j) *}} \log p\left(\mathbf{y} \mid \mathbf{i}_t^{(j)}\right)$ using (\ref{scoreresult})
    \STATE \textbf{Corrector}
    \STATE $\nabla_{\mathbf{i}_t^{(j) *}} \log p\left(\mathbf{i}_t^{(j)} \mid \mathbf{y}\right) \leftarrow \mathbf{s}_{\hat\theta}\left(\mathbf{i}_t^{(j)}, t\right) +\nabla_{\mathbf{i}_t^{(j) *}} \log p\left(\mathbf{y} \mid \mathbf{i}_t^{(j)}\right)$
    \STATE $\mathbf{z}_{\mathbf{i}} \sim \mathcal{C N}\left(\mathbf{0}, 2 \mathbf{E}_{N}\right)$
    \STATE $\epsilon_t \leftarrow \|\mathbf{z}_{\mathbf{i}}\|_2^2 /\left\|\nabla_{\mathbf{i}_t^{(j) *}} \log p\left(\mathbf{i}^{(j)}_t \mid \mathbf{y}\right)\right\|_2^2$
    \STATE $\mathbf{i}_t^{(j)} \leftarrow \mathbf{i}_t^{(j)} + \epsilon_t \nabla_{\mathbf{i}_t^{(j) *}} \log p\left(\mathbf{i}_t^{(j)} \mid \mathbf{y}\right) + \sqrt{2 \epsilon_t} \mathbf{z}_{\mathbf{i}}$
    \STATE Calculate $\nabla_{\mathbf{i}_t^{(j) *}} \log p\left(\mathbf{y} \mid \mathbf{i}_t^{(j)}\right)$ using (\ref{scoreresult})
    \STATE \textbf{Predictor}
    \STATE $\nabla_{\mathbf{i}_t^{(j) *}} \log p\left(\mathbf{i}^{(j)}_t \mid \mathbf{y}\right) \leftarrow \mathbf{s}_{\hat\theta}\left(\mathbf{i}_t^{(j)}, t\right) +\nabla_{\mathbf{i}_t^{(j) *}} \log p\left(\mathbf{y} \mid \mathbf{i}_t^{(j)}\right)$
    \STATE $\mathbf{z}_{\mathbf{i}} \sim \mathcal{C N}\left(\mathbf{0}, 2 \mathbf{E}_{N}\right)$
    \STATE $\mathbf{i}_{t-\text{d}t}^{(j)} \leftarrow \mathbf{i}_{t}^{(j)} +\left(-f(t) \mathbf{i}_t^{(j)} + g^2(t) \nabla_{\mathbf{i}_t^{(j) *}} \log p\left(\mathbf{i}^{(j)}_t \mid \mathbf{y}\right)\right)\text{d}t + g(t)\sqrt{ \text{d}t} \mathbf{z}_{\mathbf{i}}$
    \STATE \textbf{Posterior Inference}
    \STATE Calculate $\boldsymbol{\mu}_{\text{post},t}$ using (\ref{mupost})
    \STATE Calculate $\boldsymbol{\Sigma}_{\text{post},t}$ using (\ref{sigmapost})
    \STATE \textbf{Update Hyperparameter}
    \STATE ${\boldsymbol{\sigma}}^2_{t-\text{d}t} = \boldsymbol{\Sigma}_{\text{post},t}+\left|\boldsymbol{\mu}_{\text{post},t}\right|^{\odot 2}$
    \STATE \textbf{Noise Variance Estimation}
    \STATE ${\sigma}^2_{w,t-\text{d}t}=\min\left(\frac{1}{N}\left(\frac{1}{\tilde{J}}\left(\sum_{\tilde{j}=1}^{\tilde{J}}\|\mathbf{y}-\mathbf{A} \boldsymbol{\mu}_{\text{post},t} -\tilde{\mathbf{i}}_t^{(\tilde{j})}\|^2_2\right) + \operatorname{tr}(\mathbf{A} \boldsymbol{\Sigma}_{\text{post},t}\mathbf{A}^{\rm H})\right)-\frac{2b_t^2}{\beta_t^2},\zeta\right)$
    \ENDFOR
    \STATE \textbf{Threshold Detector}
    \STATE $\hat{\mathbf{r}} = \text{Detector}(\boldsymbol{\mu}_{\text{post},0})$
    \STATE \textbf{return} $\hat{\mathbf{r}}$
\end{algorithmic}
\end{algorithm}

\section{Numerical Results}\label{num}
In this section, substantial numerical experiments are carried out to demonstrate the effectiveness and robustness of the proposed DMDD algorithm. Firstly, the ability of the score-based neural network in capturing the key characteristics of structured jamming is verified. Secondly, the target detection performance of the DMDD method in the presence of structured jamming is assessed. Thirdly, performances between the DMDD algorithm and benchmark methods are provided under various SNR conditions. Finally, real clutter data is employed to validate the effectiveness of the DMDD method. Experiments were performed on a system with an NVIDIA RTX 4070 Ti SUPER GPU and an Intel i7-14700KF CPU.

In the following analysis, the signal-to-jamming ratio (SJR) and SNR are defined as
\begin{align}
    \text{SJR} = 10 \log_{10} \frac{\|\mathbf{A} {\mathbf{x}}\|_2^2}{\|\mathbf{i}\|_2^2},\ \text{SNR}_k = 10 \log_{10} \frac{|{x}_k|_2^2}{\sigma_{w}^2},
\end{align}
respectively, where $x_k$ is defined in (\ref{signalmodel}).

\subsection{Validation of the Score-Based Model in Jamming Learning}\label{Exppretrain}

This subsection presents an experiment to validate the ability of the score-based model in learning and characterizing typical jamming patterns.

In this experiment, the structured jamming is considered following a comb-spectrum pattern, which is a widely used electronic countermeasure against modern radar systems. The jammer transmits a set of continuous-wave (CW) signals whose frequencies are randomly distributed across the radar’s operating bandwidth. Mathematically, after down conversion and LPF, the baseband jamming signal can be expressed as
\begin{align}
    \mathbf i(t) = \sum_{k=1}^{K} A_k {\rm e}^{\text{j} 2\pi(f_0 + k\Delta f) t},
\end{align}
where $K$ is the number of CW signals, $A_k$ denotes the amplitude of the $k$th signal, $f_0$ is the starting frequency, and $\Delta f$ is the frequency spacing.
In the frequency domain, comb-spectrum jamming concentrates energy at discrete tones given by $f_k = f_0 + k\Delta f,  k = 1, 2, \dots, K$, forming a periodic spectral structure. This disrupts the correlation between the received echo and reference waveform, spreading sidelobes and periodic false alarms in the pulse compression (PC)  output \cite{PC}.

\newcommand{\tabincell}[2]{\begin{tabular}{@{}#1@{}}#2\end{tabular}}
\begin{table}[!t]
\centering
\scriptsize
\caption{Parameters Setting of Experiment 1}
\label{Exp1parameters}
\begin{tabular}{ll}
    \\[-2mm]
    \hline
    \hline\\[-2mm]
    { \small Signal Parameters}&\qquad {\small Value}\\
    \hline
    \vspace{1mm}\\[-3mm]
    Carrier frequency $f_c$   &   $8.11$ GHz\\
    \vspace{1mm}
    Frequency modulation slope $\mu$          &  $1.5$ MHz/$\mu$s\\
    \vspace{1mm}
    Pulse width $T_{\rm p}$         &  $10 {\rm \mu s}$\\
    \vspace{1mm}
    Pulse duration          &  $120 {\rm \mu s}$\\
    \vspace{1mm}
    Bandwidth $B$          &  $15$ MHz\\
    \vspace{1mm}
    Sampling frequency $f_s = 1 / T_s$         &  $31.25$ MHz\\
    \hline
    \hline
    \\[-2mm]
    { \small Jamming Parameters}&\qquad {\small Value}\\
    \hline
	\vspace{1mm}\\[-3mm]
	Number of CW signals  $[K_{\rm min},K_{\rm max}]$ & $[5,10]$ \\
    \vspace{1mm}
	Range of frequencies $[f_{\rm min},f_{\rm max}]$ & $[-7.5,7.5]$ MHz \\
	\vspace{1mm}
	Range of frequency spacing $[\Delta f_{\rm min},\Delta f_{\rm max}]$ & $[0.5,1.5]$ MHz \\
    [1mm]
    \hline
    \hline
\end{tabular}
\end{table}



To estimate the jamming, a dataset of comb-spectrum jamming signals is used to train a neural network via score matching (\ref{Iscorematch}). 
Specifically, the amplitude of each tone is drawn as
$A_k \sim \mathcal{U}(A_{\min}, A_{\max})$, where $\mathcal{U}(\cdot,\cdot)$ denotes the uniform distribution, and $[A_{\min}, A_{\max}]$ defines the feasible amplitude range.
The number of jamming sources is drawn from
$K \sim \mathcal{U}_{\text{int}}(K_{\min}, K_{\max})$,
where $\mathcal{U}_{\text{int}}(\cdot,\cdot)$ represents a discrete uniform distribution over integers.
The starting frequency of the comb-spectrum jamming follows
$f_0 \sim \mathcal{U}(f_{\min}, f_{\max} - (K-1)\Delta f)$, ensuring that all tones lie within the radar signal bandwidth $[f_{\min}, f_{\max}]$.
The frequency spacing between adjacent tones is randomly assigned as
$\Delta f \sim \mathcal{U}(\Delta f_{\min}, \Delta f_{\max})$.

The dataset containing $10^7$ jamming-only samples generated according to the parameters in Table \ref{Exp1parameters}. Training was performed with a batch size of 128 using the lightweight U-net with a learning rate of $10^{-4}$, and typically converged within $10^3$ epochs. And the pretraining process required approximately $74$ hours to complete.

\begin{figure*}
    \centering
    \subfigure[]{
    \label{timedomain_comb}
    \includegraphics[width = 2in]{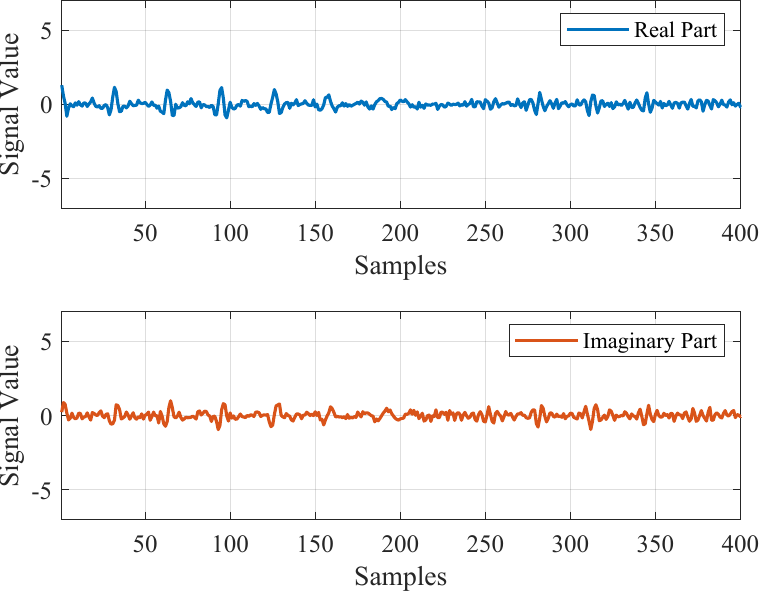}}
    \subfigure[]{
    \label{fredomain_comb}
    \includegraphics[width = 2in]{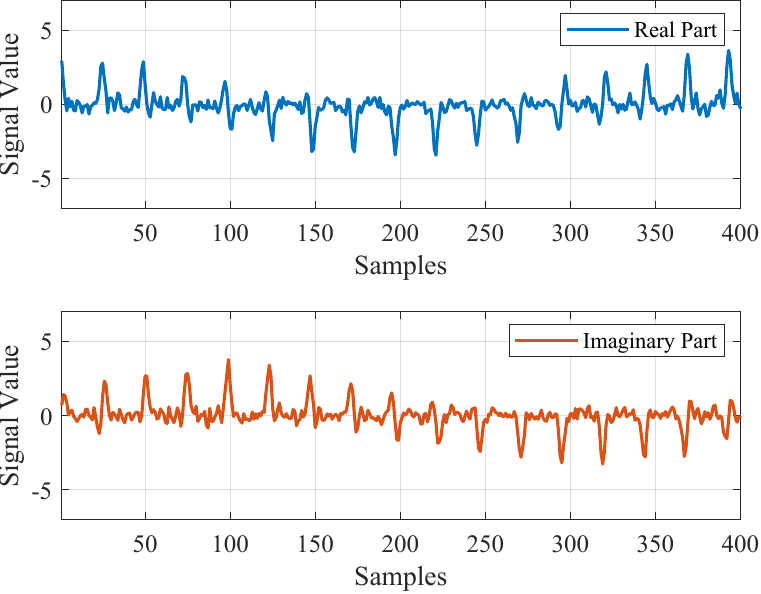}}
    \subfigure[]{
    \label{fredomain_comb}
    \includegraphics[width = 2in]{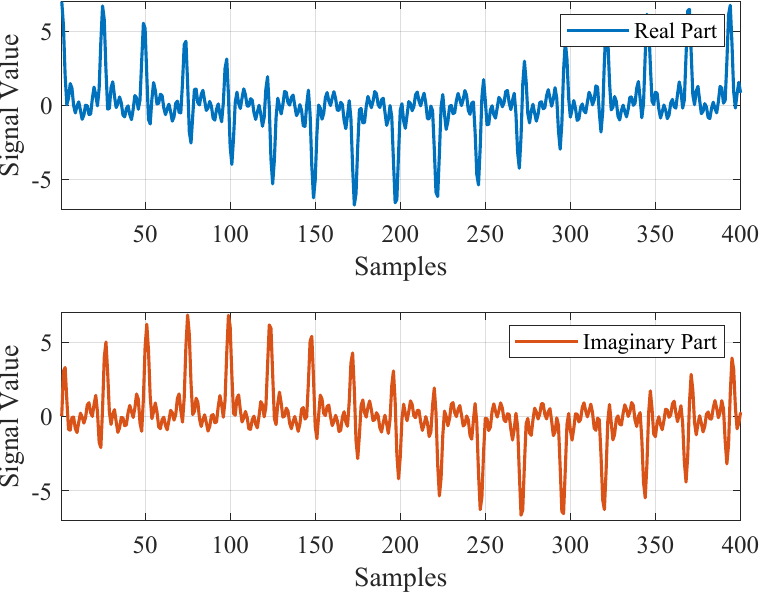}}
	\caption{Generated realizations of comb-spectrum jamming using the score at three diffusion times. (a) $t=1$; (b) $t=0.5$; (c) $t=0$.}\label{jam_gen}
\end{figure*}

\begin{figure*}
    \centering
    \subfigure[]{
    \label{freq_gen_t1}
    \includegraphics[width = 2in]{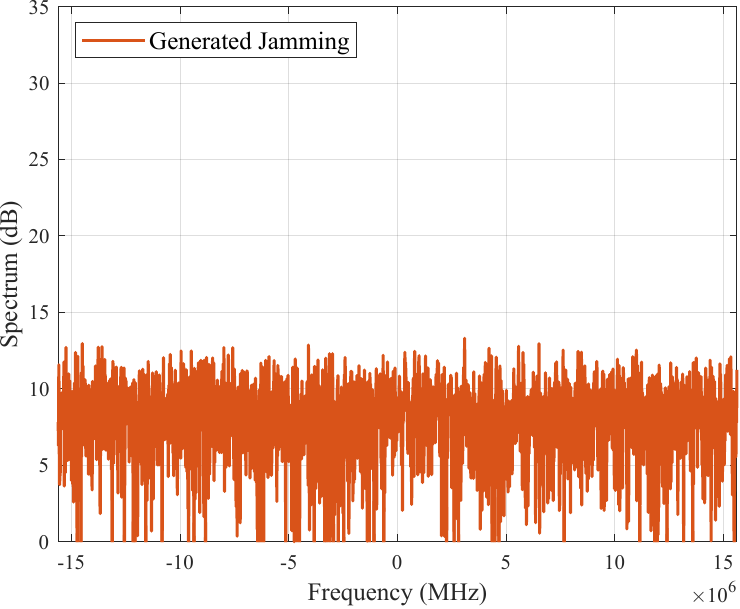}}
    \subfigure[]{
    \label{freq_gen_t05}
    \includegraphics[width = 2in]{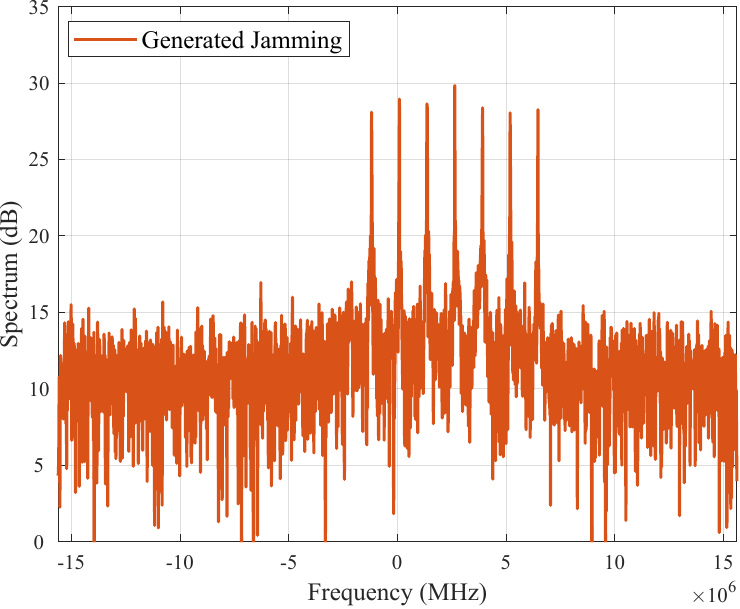}}
    \subfigure[]{
    \label{freq_gen_t0}
    \includegraphics[width = 2in]{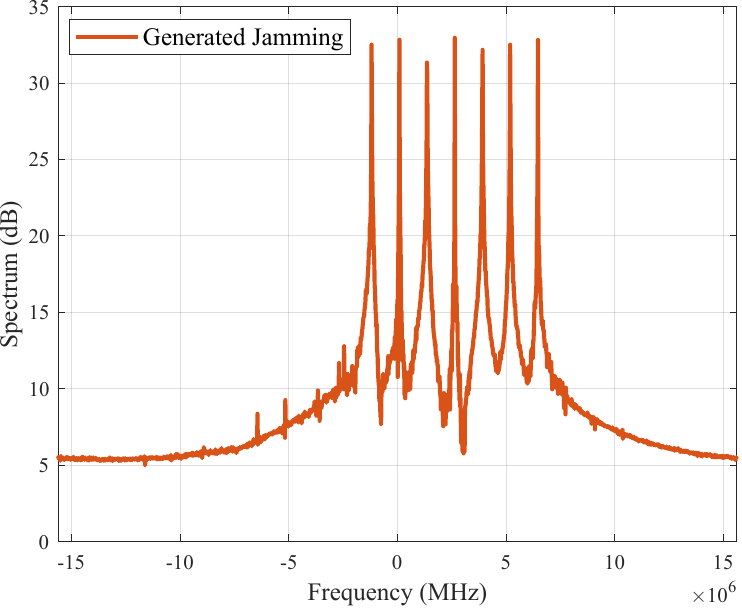}}
	\caption{Spectrum of the generated comb-spectrum jamming three diffusion times. (a) $t=1$; (b) $t=0.5$; (c) $t=0$.}\label{jam_gen_freq}
\end{figure*}

Using the pretrained score network, jamming samples can be synthesized directly by running the reverse diffusion process conditioned only on the learned score. Fig. \ref{jam_gen} shows generated realizations at three diffusion times: $t=1$, $t=0.5$ and $t=0$. The sequence illustrates progressive refinement of the waveform, and the spectrum of the generated jamming are presented in Fig. \ref{jam_gen_freq}. As shown in Fig. \ref{freq_gen_t1}, the generated signal is pure noise at $t = 1$. From Fig. \ref{freq_gen_t05}, it can be observed that at $t = 0.5$, the spectral characteristics of the comb-spectrum signal begin to emerge. Fig. \ref{freq_gen_t0} presents the spectrum corresponding to the end of the diffusion process at $t = 0$, and it can be seen that the pre-trained score model generates a comb-like jamming with starting frequency $f_0=-1.19$ MHz, frequency spacing $\Delta f=1.09$ MHz and the number of jamming sources $K=7$. These results indicate that the score model has successfully captured the structure of the comb-spectrum jamming and can generate such jamming waveforms, demonstrating the score model’s effectiveness in learning the structure of comb-spectrum jamming.

\subsection{Effectiveness of the DMDD Method}\label{sub2}
In this subsection, effectiveness of the proposed DMDD algorithm is investigated under both on-grid and off-grid cases. We conduct the conventional PC method as a baseline for comparison. In each cases, the presence of strong comb-spectrum jamming poses a significant challenge to conventional algorithm. We also compare the DM-SBL method in \cite{yifan} under each cases, which constructs two independent diffusion processes to perform joint posterior sampling.

Consider a scenario with two targets, and the SNRs of the two targets are $\rm SNR_1 = \rm SNR_2 = -5$ dB. Note that after coherent integration, the integrated SNR reaches $19.96$ dB. The SJR is set to $-20$ dB. The signal and jamming parameters are consistent with those summarized in Table \ref{Exp1parameters}. The grid spacing for the range bin set $\mathcal{R}$ is set $4.8$ m. The threshold parameter $T_h$ is set $16.8$ dB in the jamming scenario. For DM-SBL, we also apply the same constant threshold detector and the noise variance $\sigma_w^2$ is treated to be known.
\subsubsection{The on-grid Case}
\begin{figure*}
    \centering
    \subfigure[]{
    \label{PC_1pul}
    \includegraphics[width = 2in]{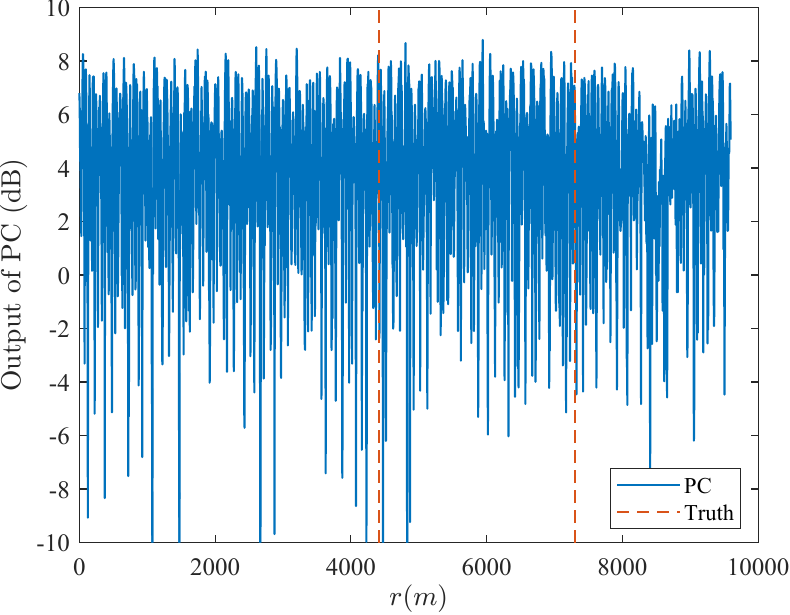}}
    \subfigure[]{
    \label{DM_1pul}
    \includegraphics[width = 2in]{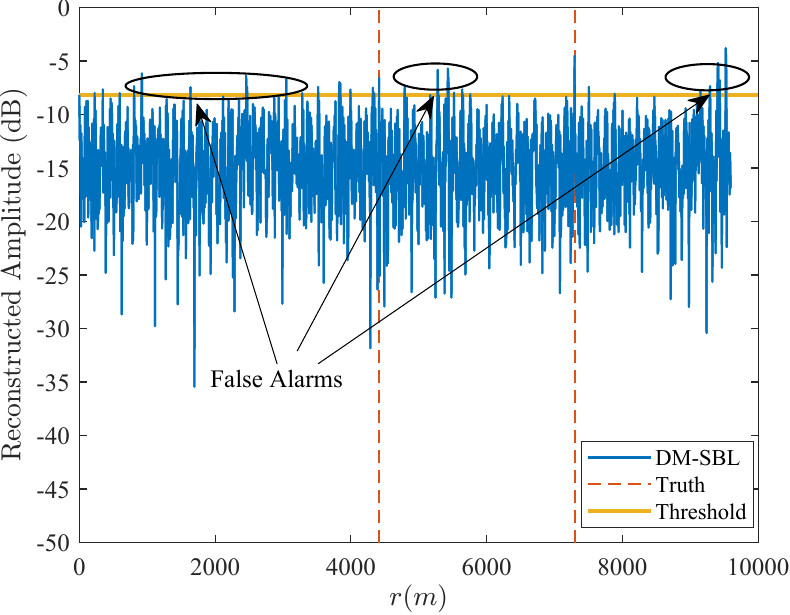}}
    \subfigure[]{
    \label{DMDD_1pul}
    \includegraphics[width = 2in]{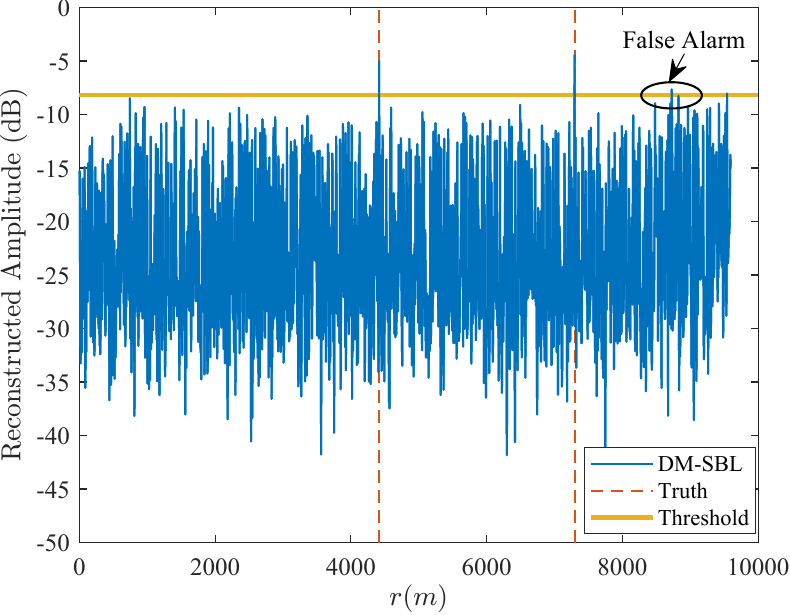}}
	\caption{On-grid case. Output of the PC method, the DM-SBL method and the proposed DMDD method: (a) PC; (b) DM-SBL; (c) DMDD. The red dotted line represents the true ranges of the targets and the black ellipses denote the false alarms.}\label{on_pul_1}
\end{figure*}
Target $1$ is located at range $r_1 = 4416$ m at the $921$th range grid, and target $2$ is located at $r_2 = 7296$ m at the $1521$th range grid.
In Fig. \ref{PC_1pul}, the output of PC is presented. It can be seen that the two targets are completely masked by strong jamming, with numerous false alarms scattered across the range bins. In such jamming scenarios, the conventional method is unable to detect the targets. The reconstructed amplitude obtained by DM-SBL is provided in Fig. \ref{DM_1pul}. After applying the threshold detector, the range estimates are obtained. Results show that DM-SBL successfully detects both targets, with the estimated amplitude of target $1$ exceeding the detection threshold of $1.58$ dB and the estimated amplitude of target $2$ exceeding the detection threshold of $3.63$ dB. However, it is also noticeable that DM-SBL produces a relatively high number of false alarms.

The reconstructed amplitude obtained by DMDD is provided in Fig. \ref{DMDD_1pul}. Compared to DM-SBL, DMDD suppresses the jamming more effectively. After applying the threshold detector, the range estimates are obtained. Results show that DMDD also successfully detects both targets, with the estimated amplitude of target 1 exceeding the detection threshold of $3.14$ dB and the estimated amplitude of target 2 exceeding the detection threshold of $3.72$ dB. Notably, only one false alarm was observed when using DMDD in this experiment, indicating that it provides better suppression of false alarms. In terms of runtime, the DM-SBL method requires $26.27$ seconds to reach a stable solution, whereas the proposed DMDD method requires $11.73$ seconds. DM-SBL method require a longer runtime due to multiple diffusion steps to sample $\tilde{\mathbf{x}}$, and the DMDD algorithm achieves higher efficiency by directly generating posterior samples. At present, the computational efficiency is insufficient for real-time implementation. Future work will focus on further reducing algorithmic complexity to enable real-time processing.

\subsubsection{The off-grid Case}
Target $1$ is located at range $r_1 = 4418.4$ m, offset by $2.4$ m from the nearest range grid with index $921$, and target $2$ is located at $r_2 = 7298.4$ m, also offset by $2.4$ m from the nearest range grid with index $1521$. Notably, the ranges of the two targets fall exactly midway between adjacent discretized grid points, representing a typical off-grid scenario.

\begin{figure*}
    \centering
    \subfigure[]{
    \label{off_grid_PC}
    \includegraphics[width = 2in]{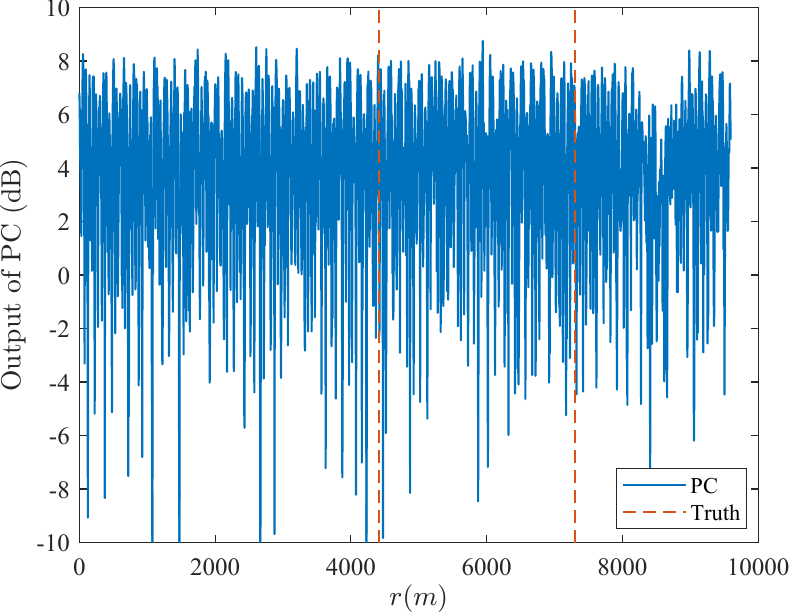}}
    \subfigure[]{
    \label{off_grid_DM}
    \includegraphics[width = 2in]{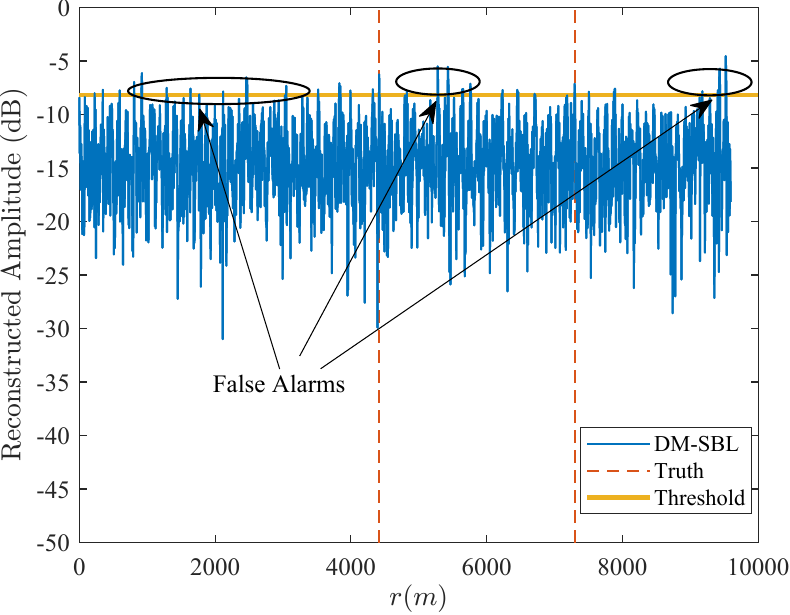}}
    \subfigure[]{
    \label{off_grid_PS}
    \includegraphics[width = 2in]{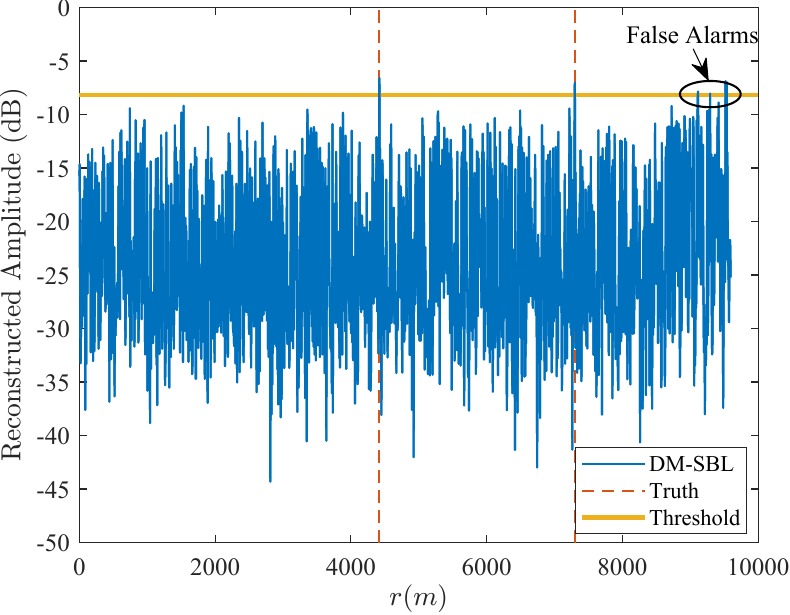}}
	\caption{Off-grid case. Output of the PC method, the DM-SBL method and the proposed DMDD method: (a) PC; (b) DM-SBL; (c) DMDD. The red dotted line represents the true ranges of the targets and the black ellipses denote the false alarms.}\label{on_pul_1}
\end{figure*}
In Fig. \ref{off_grid_PC}, the output of PC is shown, demonstrating that it fails to detect either target due to the overwhelming interference caused by strong jamming.
The reconstructed amplitude obtained by DM-SBL is provided in Fig. \ref{off_grid_DM}. It can be seen that the DM-SBL method can detect target $1$ with the estimated amplitude exceeding the detection threshold of $1.91$ dB and target $2$ with the estimated amplitude exceeding the detection threshold of $1.02$ dB. Similar to the on-grid case, DM-SBL also introduces much false alarms. The reconstructed amplitude obtained by DMDD is provided  in Fig. \ref{off_grid_PS}. It can be observed that the proposed DMDD method remains robust under the off-grid conditions, accurately detecting both targets to the nearest grid points to their true ranges, with the estimated amplitude of target $1$ exceeding the detection threshold of $1.55$ dB and the estimated amplitude of target $2$ exceeding the detection threshold of $1.15$ dB. These results demonstrate the superior capability of the proposed DMDD to suppress CAstructured jamming and detect targets.

\subsection{Performance Comparisons}
In this subsection, how the detection probability $P_D$ and false alarm probability $P_{FA}$ of the algorithms vary with the integrated SNR under jamming conditions are investigated. To provide the evaluation, the DMDD algorithm is compared against DM-SBL \cite{yifan} and the following benchmark algorithms:
\begin{itemize}
    \item SBL \cite{SBL}: The jammings are ignored, and the prior of $\tilde{\mathbf{x}}$ is estimated using a standard SBL framework. 
    
    \item SBL-Second Order Moment (SBL-SOM): The jammings are assumed to be Gaussian distributed, and second-order moment information is leveraged for signal recovery.

    \item Alternating Direction Method of Multipliers (ADMM) \cite{ADMM}: The estimation of $\tilde{\mathbf{x}}$ is formulated as a sparsity-constrained optimization problem, which is solved iteratively using ADMM by assuming the jammings to be sparse in the frequency domain. 
    Let $\mathbf{a}(f) = [1, e^{\mathrm{j}2\pi f}, \cdots, e^{\mathrm{j}2\pi f(N-1)}]^{\mathrm{T}}$ represent a complex sinusoid of frequency $f$, and define the overcomplete dictionary as $\mathbf{F}(\mathbf{f}) = [\mathbf{a}(f_1), \cdots, \mathbf{a}(f_M)]$, where the frequency bin set $\mathcal{F} = \{f_m\}_{m=1}^M$ consists of $M = 4N$ uniformly spaced grid points distributed between $f_{\min}$ and $f_{\max}$.
    And the receive signal is modeled as 
    \begin{align}
        \mathbf{y}=\mathbf{A}(\tilde{\mathbf{r}})\tilde{\mathbf{x}} + \mathbf{F}(\mathbf{f})\mathbf{z}+\mathbf{w},
    \end{align}
    where $\mathbf{z}$ is the amplitude vector of the jamming. And the corresponding optimization problem is
    \begin{align}
    \hat{\tilde{\mathbf{x}}},\hat{\mathbf{z}}= \underset{\tilde{\mathbf{x}} \in \mathbb{C}^N,\ \mathbf{z}\in \mathbb{C}^M}{\text{argmin}}\left(\|\mathbf{y}-\mathbf{A} \tilde{\mathbf{x}}-\mathbf{Fz}\|_2^2+\lambda\left(\|\tilde{\mathbf{x}}\|_1+\|\mathbf{z}\|_1\right)\right),
    \end{align}
    where the regularization parameter $\lambda>0$ admits balancing the data fitting fidelity versus the sparsity level in $\tilde{\mathbf{x}}$ and $\mathbf{z}$.
\end{itemize}

The radar and jamming parameters used in this experiment are consistent with those listed in Table \ref{Exp1parameters}. The grid spacing and threshold parameters setting follow the configuration in Subsection \ref{sub2}. For each trial, six targets are randomly placed in the fast time domain, and the integrated SNRs of targets vary from $12$ dB to $22$ dB. The SJR is set $–20$ dB, and $500$ Monte Carlo trials are conducted.

\begin{figure*}
    \centering
    \subfigure[]{
    \label{pfa_comparation}
    \includegraphics[width = 2.6in]{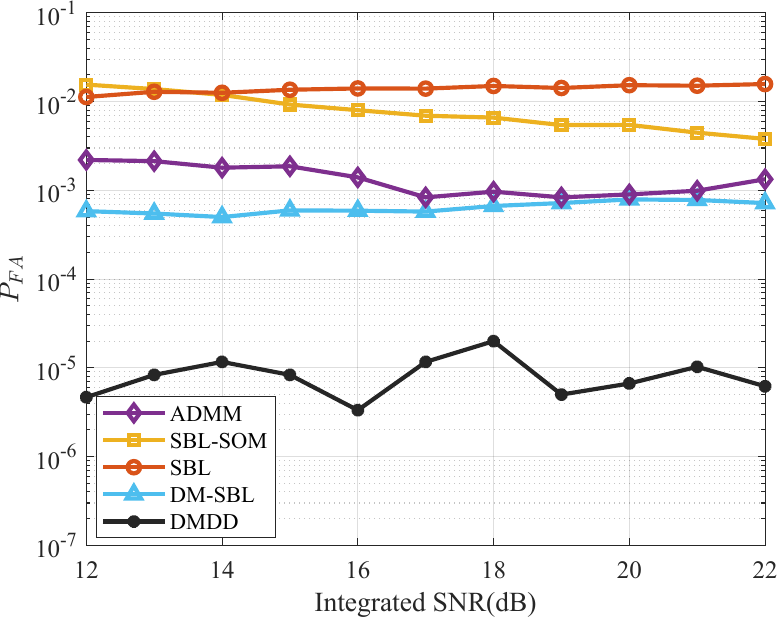}}
    \subfigure[]{
    \label{pd_comparation}
    \includegraphics[width = 2.6in]{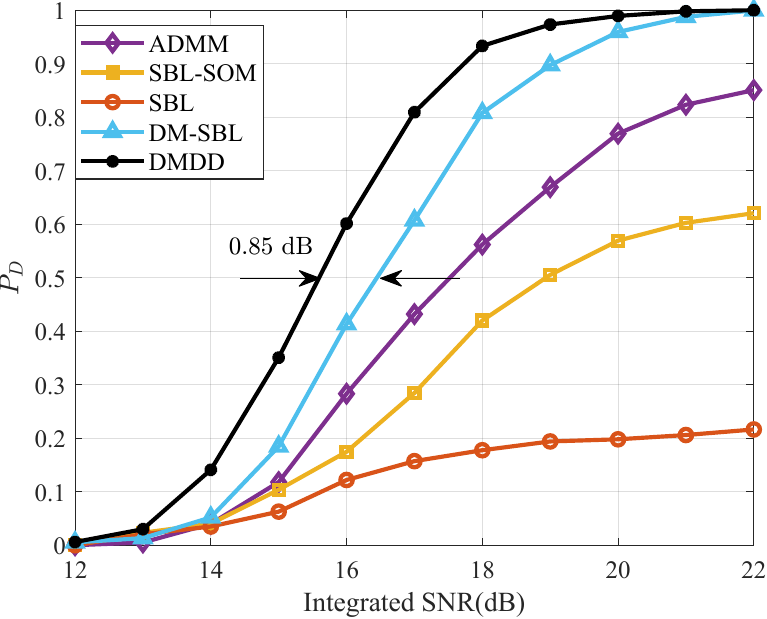}}
	\caption{Performance comparisons between different algorithms with SIR = $-20$ dB. (a) False alarm probability $P_{FA}$ versus integrated SNR; (b) Detection probability $P_D$ versus integrated SNR.}\label{comparation}
\end{figure*}

The false alarm probability of the methods versus integrated SNR are shown in Fig. \ref{pfa_comparation}. When the integrated SNR increases from $12$ dB to $22$ dB, the false alarm probabilities of the SBL and SBL-SOM methods remain around $10^{-2}$, with their maximum $P_{FA}$ being $1.57\times10^{-2}$ and $1.55\times10^{-2}$, respectively. The ADMM and DM-SBL methods achieve lower false alarm probabilities of approximately $10^{-3}$, with corresponding maximum $P_{FA}$ being $2.20\times10^{-3}$ and $7.89\times10^{-4}$, respectively. In contrast, the proposed DMDD method attains the lowest false alarm probability of about $10^{-5}$, with the maximum $P_{FA}$ being $1.99\times10^{-5}$. These results indicate that DMDD provides the most effective suppression of false alarms.

From Fig. \ref{pd_comparation}, it can be observed that all methods exhibit monotonically increasing detection probability as the integrated SNR increases. When the integrated SNR increases from $12$ dB to $22$ dB, the detection probabilities of the SBL, SBL-SOM, ADMM, DM-SBL, and DMDD methods increase from nearly zero to $0.22$, $0.62$, $0.85$, $1.0$ and $1.0$, respectively. Among these algorithms, DMDD achieves the best detection performance, followed by DM-SBL, ADMM, SBL-SOM, and SBL. When the detection probability reaches 0.5, the corresponding integrated SNRs of DMDD and DM-SBL are approximately $15.57$ dB and $16.42$ dB, respectively. And the corresponding false alarm probability $8.33\times10^{-6}$ of DMDD is two orders of magnitude lower than that of DM-SBL, which is $5.89\times10^{-4}$, yielding a detection performance gain of at least $0.85$ dB.

\subsection{Real-clutter Dataset with Injected Synthetic Targets}\label{real}

In this subsection, real radar data collected from a field experiment are processed to validate the effectiveness of the proposed DMDD method. The radar parameters used in the experiment are the same as listed in Table \ref{Exp1parameters}. The test scenario includes strong clutters consisting of ground and building clutter. Although we focus on the jamming scenario, the proposed DMDD suppresses the clutter and perform target detection effectively. Therefore, we learn the clutter and $\mathbf{i}$ is modeled as the clutter. We show the ability of clutter suppression in the following experiments.


\begin{figure*}
    \centering
    \subfigure[]{
    \label{real_PC_1pul}
    \includegraphics[width = 2.6in]{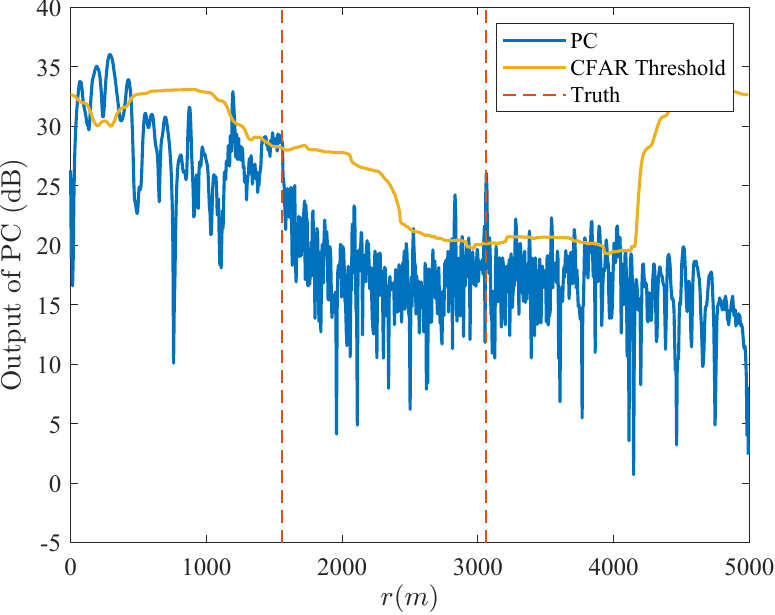}}
    \subfigure[]{
    \label{real_DMDD_1pul}
    \includegraphics[width = 2.6in]{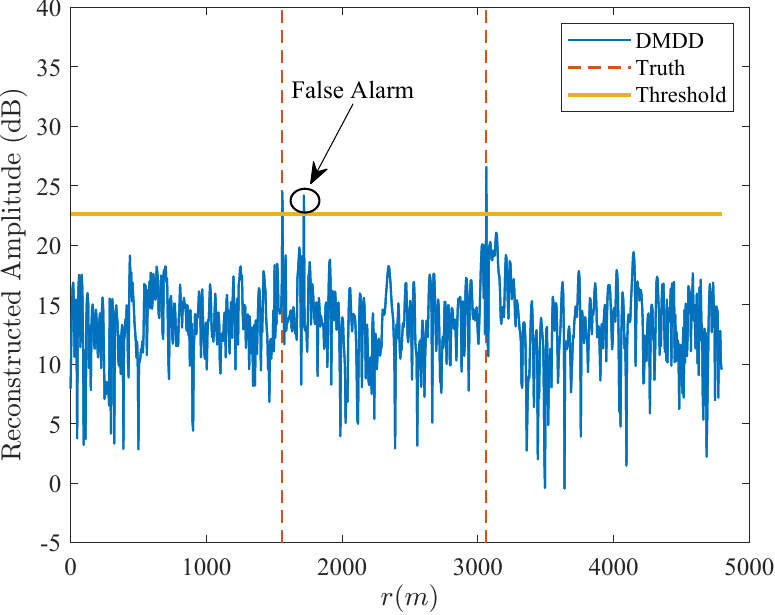}}
	\caption{Output of PC and DMDD method in real data experiment. (a) PC. (b) DMDD. The red dotted line represents the true ranges of the targets and the black ellipses denote the false alarms.}\label{real_data_1pul}
\end{figure*}
Echo dataset containing $1\times10^6$ pulses was collected under environments dominated by clutters and noise. The dataset is randomly divided into training and testing subsets with a ratio of $80/20$, where the training set is employed to pre-train the jamming score and the testing set is reserved for performance evaluation. The parameters of the lightweight U-net is the same as used in Subsection \ref{Exppretrain}, and the pretraining process required approximately $37$ hours to complete.

\subsubsection{Experiment $1$}
For validation, a semi-synthetic radar data is constructed by injecting two synthetic target returns into the real-clutter data of the testing set. Target $1$ is located at $1561$ m and target $2$ is located at $3064$ m. Amplitudes of each targets are set $28$ dB. The output of PC is shown in Fig. \ref{real_PC_1pul}, where significant clutter is observed within ranges below $1500$ m. And a standard cell-averaging constant false alarm rate (CA-CFAR) detector, with the false alarm rate being set $1\times10^{-5}$, is applied to the output of PC. 
As shown in Fig. \ref{real_PC_1pul}, the PC method is able to detect target $1$; however, a dense cluster of false alarms appears in its immediate vicinity, making it difficult to reliably distinguish the true target. Target $2$ lies in a region with weaker clutters, enabling higher-quality detection, while a considerable number of false alarms are still present nearby.

The reconstructed amplitude obtained using the proposed DMDD method is presented in Fig. \ref{real_DMDD_1pul}. The noise variance $\sigma_w^2$ is estimated to be $902.69$, and the threshold parameter $T_h$ is set $17.9$ dB. It can be seen that the DMDD method successfully detects the two targets at their nearest range bins, namely $1560$ m and $3062.4$ m, with the estimated amplitude of target $1$ exceeding the detection threshold of $1.86$ dB and the estimated amplitude of target $2$ exceeding the detection threshold of $3.68$ dB. It can be seen that only a single false alarm appears near target $1$, and DMDD effectively suppresses clutter distributed within ranges below $1500$ m.

\subsubsection{Experiment $2$}
\begin{figure*}
    \centering
    \subfigure[]{
    \label{Pf_comparasion_real}
    \includegraphics[width = 2.6in]{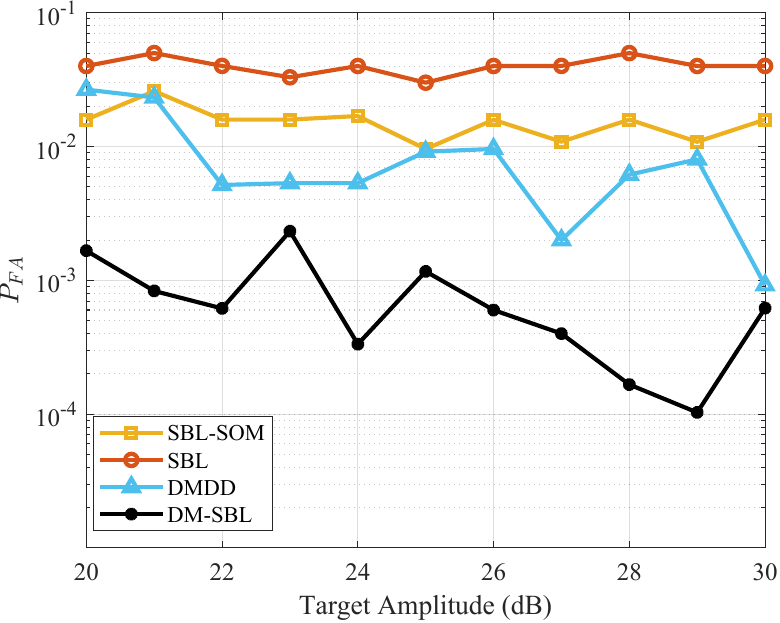}}
    \subfigure[]{
    \label{Pd_comparasion_real}
    \includegraphics[width = 2.6in]{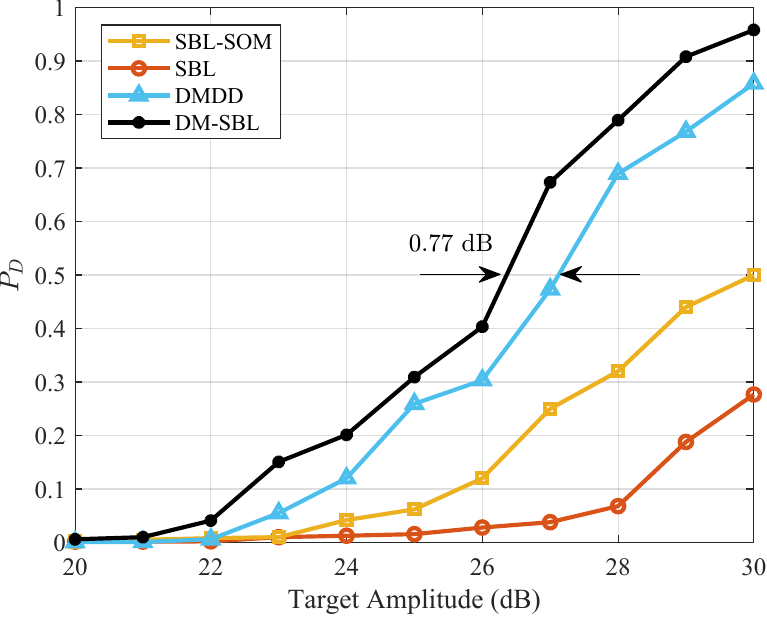}}
	\caption{Performance comparisons between different algorithms using real-clutter dataset. (a) False alarm probability $P_{FA}$ versus target amplitude; (b) Detection probability $P_D$ versus target amplitude.} \label{comparasion_real}
\end{figure*}
To further evaluate detection performance under real-clutter scenarios, 500 Monte Carlo trials using the testing dataset is conducted. In each trial, six targets with identical amplitudes ranging from $20$ dB to $30$ dB are injected into the raw data, and the false alarm probability and detection probability of different methods were statistically assessed. 

The false alarm probability of the methods versus target amplitude are shown in Fig. \ref{Pf_comparasion_real}.
As the target amplitude increases from $20$ dB to $30$ dB, the SBL and SBL-SOM methods exhibit the highest false alarm probabilities, with their maximum values being $5.01\times10^{-2}$ and $2.60\times10^{-2}$, respectively. The DM-SBL method yields moderately lower false alarm probability, with a maximum probability of $2.67\times10^{-2}$. In contrast, the proposed DMDD method achieves the lowest false alarm probability among all methods, with its maximum value being $2.33\times10^{-3}$.

From Fig. \ref{Pd_comparasion_real}, it can be observed that the detection probabilities of all methods increase with the target amplitudes. As the target amplitudes increases from $20$ dB to $30$ dB, the detection probabilities rise from nearly zero to $0.28$ for SBL, $0.50$ for SBL-SOM, $0.86$ for DM-SBL, and $0.96$ for DMDD. Among these approaches, DMDD exhibits the best overall detection performance, followed by DM-SBL, SBL-SOM, and SBL. When the detection probability reaches $0.5$, the corresponding target amplitude is approximately $27$ dB for DM-SBL and DMDD. At this operating point, the false alarm probability of DMDD is $4.00\times10^{-3}$, which is lower than that of DM-SBL, which is $2.00\times10^{-2}$, yielding a detection performance gain of at least $0.77$ dB.

These experiments verify that DMDD maintains robustness and detection reliability when applied to real-clutter radar data.

\section{Conclusion}\label{con}
This paper proposes a Diffusion-based Model and Data Dual-driven (DMDD) algorithm for multitargets detection and estimation in the presence of structured mainlobe jamming. The approach leverages a score-based diffusion model to learn and sample from complex jamming distributions, while simultaneously incorporating a model-based SBL strategy to estimate target amplitudes. Unlike previous diffusion-based methods requiring iterative sampling for both signal and jamming, the proposed DMDD method constructs only a single diffusion process for jamming, allowing for more efficient posterior inference of the target. The DMDD method is highly flexible, as it allows a pretrained score model of the jamming to be naturally incorporated into the model-driven component, and then posterior inference can be carried out. Extensive simulation results demonstrate that DMDD can suppress false alarms more effectively and exhibits better detection performance. The DMDD method offers a promising solution for radar target detection in challenging operating environments.

\end{document}